# Photo de-mixing in Dion-Jacobson two-dimensional mixed halide perovskites

Ya-Ru Wang,[1] Alessandro Senocrate,[1] Marko Mladenović,[2] Algirdas Dučinskas,[1,3] Gee Yeong Kim,[1] Ursula Röthlisberger,[2] Jovana V. Milić,[3] Davide Moia,[1]* Michael Grätzel,[1,3] Joachim Maier[1]*

[1] Max Planck Institute for Solid State Research, Stuttgart, Germany.

[2] Laboratory of Computational Chemistry and Biochemistry, Institute of Chemical Sciences and Engineering, École Polytechnique Fédérale de Lausanne (EPFL), Lausanne, Switzerland.

[3] Laboratory of Photonics and Interfaces, Ecole polytechnique Fédérale de Lausanne (EPFL), Lausanne, Switzerland.

*d.moia@fkf.mpg.de; office-maier@fkf.mpg.de

**Abstract**

Two-dimensional (2D) halide perovskites feature a versatile structure, which not only enables the fine-tuning of their optoelectronic properties but also makes them appealing as model systems to investigate the fundamental properties of hybrid perovskites. In this study, we analyzed the changes in the optical absorption of 2D Dion-Jacobson mixed halide perovskite thin films (encapsulated) based on $(PDMA)Pb(I_{0.5}Br_{0.5})_4$ (PDMA: 1,4-phenylenedimethanammonium spacer) exposed to a constant illumination. We demonstrate that these 2D mixed-halide perovskites undergo photo de-mixing with direct transformation from the pristine phase to the de-mixed phases. Almost complete re-mixing of these phases occurs when the sample is left in the dark, showing that the process is reversible in terms of optical properties. On the other hand, exposure to light appears to induce structural changes in the thin film that are not reversible in the dark. We have further investigated temperature-dependent absorption measurements under light to extract the photo de-mixed compositions and to map the photo-miscibility-gap. This work thereby reveals that photo de-mixing occurs in Dion-Jacobson two-dimensional hybrid perovskites and provides strategies to address the role of light in the thermodynamic properties of these materials.

**Keywords:** Mixed halide perovskites, 2D, phase stability, photo de-mixing, photo-miscibility-gap

**Introduction**

Organic-inorganic hybrid halide perovskites show intriguing application potential for various optoelectronic devices, including solar cells,[1] light-emitting diodes,[2] field-effect transistors,[3] and photo detectors.[4] Three-dimensional (3D) perovskites have the general formula $ABX_3$ (with the A cation placed in the cavities of corner-sharing $BX_6$ octahedra network) are prime candidates as light harvesters in solar cells.[5] Two-dimensional (2D) hybrid perovskites add additional degrees of freedom to modulate the electronic and optical properties of halide perovskites with large organic cations (spacers) separating layers of corner-sharing metal halide octahedra planes.[6] Depending on the relative orientation of the inorganic slabs and the spacer, these systems can be classified into Ruddlesden-Popper (RP) ($A'_2A_{n-1}B_nX_{3n+1}$, mostly A' monovalent spacer separating the inorganic slabs with half-a-unit-cell relative offset) [7] and Dion–Jacobson (DJ) ($A''A_{n-1}B_nX_{3n+1}$, mostly A'' divalent spacer separating well-aligned inorganic slabs) phases.[8] Here, *n* represents the number of the metal halide planes between spacer planes.[9]

Apart from the nature of the organic cations, halide composition affects their properties and halide substitution or mixing provide an effective way to vary the material properties of both 2D and 3D halide perovskites.[10] However, 3D mixed halide perovskites have been shown to undergo segregation under illumination into Br-rich and I-rich compositions (*photo de-mixing*). Such segregation process appears to be largely reversible as in the dark the de-mixed phases re-mix to form the original phase (*dark re-mixing*).[11] Various studies investigated photo de-mixing in mixed halide perovskites with the materials' miscibility gap based on thermodynamic and kinetic arguments.[12] As far as thermodynamics are concerned, polaron formation and strain effects,[13] bandgap reduction due to iodide rich phase formation [14] and local electric fields effects[15] were proposed to explain its origin. In our recent work, we have invoked selective self-trapping in iodide-based and bromide-based perovskites as a possible driving force for the process.[16] The large polarizability associated with the iodide-rich environment favors self-trapping of photo-generated holes related with iodine interstitial formation. Hole accumulation achieved by contacting mixed halide perovskites with electron transport layers[17] and through electrochemical anodic bias[18] also provide direct experimental evidence for the important role of the hole concentration on the photo de-mixing behavior in 3D mixed halide perovskites. Reports on the reversible $Pb^{2+}/Pb^0$ and $I^-/I_3^-$ redox chemistry under light also gave important insights into the possible trapping mechanism.[19]

Elucidating the origin of photo de-mixing and its relation to ion transport and defect chemistry of mixed halide perovskites is of substantial conceptual importance, as it would provide guidelines for suppressing photo-induced phase instability, optimizing halide alloying, and opening up opportunities for new optoelectronic devices.[20] Evidence for suppressed ion transport in 2D halide perovskites compared with classic 3D hybrid perovskites,[21] as well as enhancement of ion transport under light in 2D systems have been suggested.[22] Recently, photo de-mixing in 2D RP mixed halide perovskites was reported, also emphasizing the role of the spacer in dictating the occurrence of the photo-induced phase segregation[23]. A number of remaining questions require clarification, including the degree of reversibility of the photo de-mixing process in 2D mixed halide perovskites and the relevance of photo de-mixing for DJ systems. More in general, the understanding of the phase behavior of 2D halide perovskites would shed light on fundamental questions related to the role of confinement in the determination of their ionic properties. These aspects have not been addressed in detail to date. The improved stability of 2D perovskites[21b] compared to their 3D counterparts also makes them suitable model systems for investigating photo de-mixing.

Here, we report on photo de-mixing and dark re-mixing for 2D Dion-Jacobson thin films $(PDMA)Pb(I_{0.5}Br_{0.5})_4$ based on 1,4-phenylenedimethanammonium ($PDMA^{2+}$) spacer. By monitoring the in-situ evolution of optical absorption, we demonstrate that encapsulated 2D mixed-halide perovskites undergo de-mixing under illumination with almost complete re-mixing in the dark. This photo de-mixing process is accompanied by an irreversible reduction in crystallinity of the thin films, as probed using X-ray diffraction. Finally, we developed and compared experimental methods that can be used to investigate the phase behavior of materials that undergo photo de-mixing. Based on the analysis of the temperature-dependent absorption spectra of films under illumination, we provide information on the photo-induced miscibility gap for 2D Dion-Jacobson mixed halide perovskites.

**Results and discussion**

2D Dion-Jacobson halide perovskite thin films based on $(PDMA)Pb(I_{1-x}Br_x)_4$ composition (**Figure 1a**) with Br content x = 0 ($(PDMA)PbI_4$), x = 0.5 ($(PDMA)Pb(I_{0.5}Br_{0.5})_4$) and x = 1 ($(PDMA)PbBr_4$) were fabricated following the procedure described in **Section S1** of the Supporting Information. All films were encapsulated with poly(methyl methacrylate) (PMMA) unless otherwise stated. **Figure 1b** shows the XRD patterns of these films. The diffraction peaks at low angles ($2\theta$ = 7.11°, 7.16°, 7.19° for x = 0, 0.5 and 1, respectively) correspond to the expected Pb-Pb interlayer distance, indicating the successful preparation of the layered structure for these thin films.[24] The increase of the bromide content results in a decrease of such distance (12.43 Å, 12.34 Å, and 12.30 Å for x = 0, 0.5 and 1, respectively), due to the smaller radius of the bromide ion compared with the iodide ion. Calculated diffraction data using density functional theory (DFT) of optimized structural models allow us to identify the observed peaks, corresponding to different orientations (as indicated in **Figure 1b**). For the mixed halide composition x = 0.5, we explored several different possible arrangements of the iodide and bromide ions in the structure (mixtures 1-4) and calculated their structural and electronic properties (see **Section S2** in the Supporting Information, **Figure S4**). We have further analyzed thin films by UV-vis absorption spectroscopy, highlighting a pronounced excitonic absorption peak for each of the films (**Figure 1c**), which can be attributed to the optical transitions in the 2D quantum wells.[25] By increasing the Br content, the wavelength associated with the excitonic peak decreases (512 nm, 450 nm, 401 nm), in accordance with the expected changes in the resulting optical properties of layered perovskite materials. [10b, 26] We have also demonstrated the successful fabrication of thin films with other intermediate compositions (see **Figure S5** in the Supporting Information).

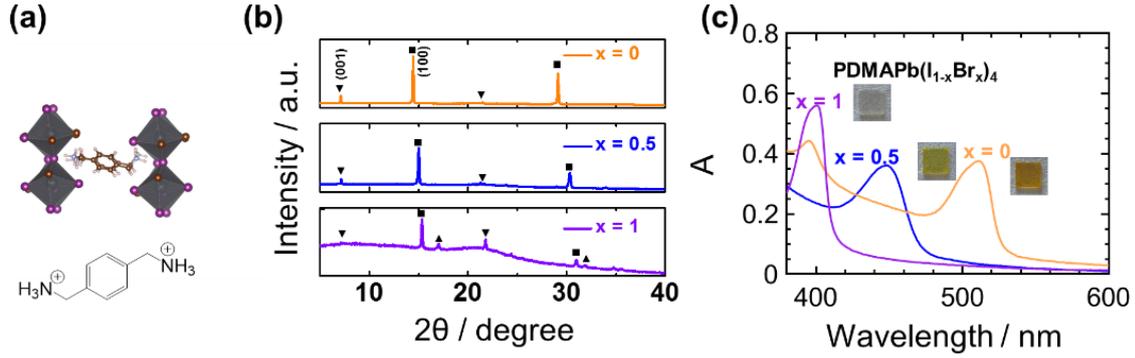

**Figure 1.** (a) Schematic representation of the (PDMA)Pb(I$_{1-x}$Br$_x$)$_4$ (n = 1) layered structure; the chemical structure of the (PDMA) spacer is shown at the bottom. (b) XRD patterns and (c) UV-Vis spectra of 2D layered perovskite thin films on quartz substrates: (PDMA)PbI$_4$ (yellow line), (PDMA)Pb(I$_{0.5}$Br$_{0.5}$)$_4$ (blue line), (PDMA)PbBr$_4$ (purple line). Photographs of relevant thin film samples with corresponding compositions are included in the inset. Diffraction peaks in (b) with ▼, ▲, and ■ representing (00l), (0k0) and (h00) Bragg planes, respectively.

Having evidenced the formation of 2D structures in thin films, we investigated their optical response under illumination. The exposure to light induced no significant change to the optical absorption spectrum of the x = 0 (iodide) and minor changes (see discussion below) for the x = 1 (bromide) 2D perovskite thin films (**Figure S6**). On the other hand, we observed a pronounced change of the absorbance for a thin film of the 50% iodide 50% bromide composition under illumination. In the following, we refer to the initial homogeneous composition of the film before exposure to light as $x_{initial}$ (in this case $x_{initial}$ = 0.5). The spectral evolution indicates a gradual phase transformation from the pristine phase (PDMA)Pb(I$_{0.5}$Br$_{0.5}$)$_4$ to Br-rich and I-rich phases (**Figure 2a**; the curves' color gradient from black to red refer to early to long time scale of illumination). Strikingly, during photo de-mixing, the absorbance of the sample remains approximately invariant for the two wavelengths $\lambda_1$ = 432 nm and $\lambda_2$ = 463 nm. The presence of these isosbestic points suggests that direct transformation from the pristine $x_{initial}$ = 0.5 composition to a second state comprising I-rich and Br-rich phases occurs. The change in absorbance ΔA (**Figure 2c**), obtained by subtracting the initial absorbance spectrum before illumination from each subsequent spectrum, clearly shows the formation of I-rich and Br-rich phases and a bleaching of the feature assigned to (PDMA)Pb(I$_{0.5}$Br$_{0.5}$)$_4$. The shape of ΔA remains approximately similar as a function of illumination time, suggesting that the phases that form at early time scales are already close in composition to the final quasi-equilibrium phases.

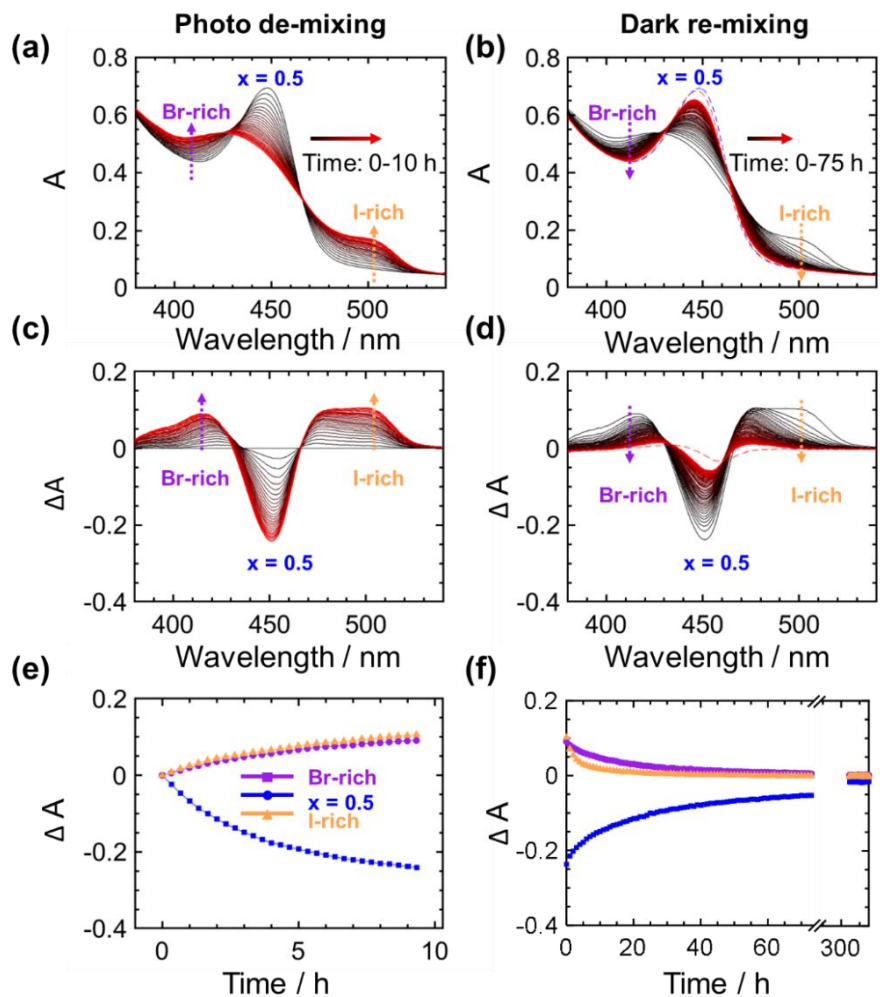

**Figure 2.** UV-Vis spectra evolution of (PDMA)Pb(I$_{0.5}$Br$_{0.5}$)$_4$ thin films (with PMMA encapsulation and in Ar atmosphere to exclude possible degradation) (a) under light (1.5 mW/cm$^2$) at 80 °C for ~10 h with 20 min interval and (b) in the dark at 80 °C for ~330 h with 1 h time interval between spectra. For the case under light, within each interval, spectra were collected by switching off the bias light for 300 s (see **Section S1** of the Supporting Information for details, **Figure S1**). The gradient from black to red denotes the time sequence. The blue dashed line represents the absorption spectra of the sample before illumination (first spectrum of panel (a)). The red dashed line corresponds to the absorbance after photo de-mixing and 330 hours of re-mixing in the dark. (c–d) The change in absorbance obtained by subtracting the reference spectrum of the pristine sample from each absorbance spectrum shown in (a) and (b), respectively. (e–f) The kinetics of the photo de-mixing and dark re-mixing processes highlighting the change in absorbance at wavelengths associated to (PDMA)Pb(I$_{0.5}$Br$_{0.5}$)$_4$ (450 nm, blue), Br-rich (414 nm, purple) and I-rich (501 nm, yellow) phases.

After photo de-mixing, illumination was removed and the sample was left in the dark. As a result, the de-mixed phases underwent re-mixing (**Figure 2b**). Only a small difference between the pristine absorption spectrum and the one after re-mixing for ~330 h in the dark (blue and red dashed lines in **Figure 2b**) was observed, indicating almost full reversibility for the sample's optical properties. The kinetics of the change in absorbance Δ*A* during photo de-mixing and dark re-mixing (displayed in **Figure 2e, f**) emphasizes the difference in time scale between the de-mixing (~10 h) and re-mixing (~300 h) processes. Analysis of XRD measurements performed on a (PDMA)Pb(I$_{0.5}$Br$_{0.5}$)$_4$ film under light at room temperature showed broadening of the peaks with a decrease in the intensity. As the XRD features did not recover in the dark, this observation points towards irreversible structural change (**Figure S27**), which may potentially be related with the formation of nano-domains after photo de-mixing (see also below). A full clarification of this process goes beyond the scope of this work, and deserves future investigation.

In order to investigate more in detail the properties of the miscibility gap for these compounds when exposed to light, which we will refer to as *photo-miscibility-gap*, we carried out temperature-dependent UV-vis absorption measurement of a thin film with $x_{initial} = 0.5$ under the same illumination conditions as shown in **Figure 2**. Starting from 100 °C, the temperature was gradually decreased to 40 °C, with the illumination intensity being kept constant. As shown in **Figure 3**, while no significant change was observed for the shape of the feature associated to the Br-rich phase, the absorption signal detected for the I-rich phase showed a significant red shift with decreasing temperature. The reversibility of the process was inspected by performing a second temperature scan back to high temperature under the same conditions (**Figure 3b, d**). Strikingly, the absorption features of the I-rich phase shifted back to the same wavelength region as originally observed for the same temperature, suggesting a reversible behavior.

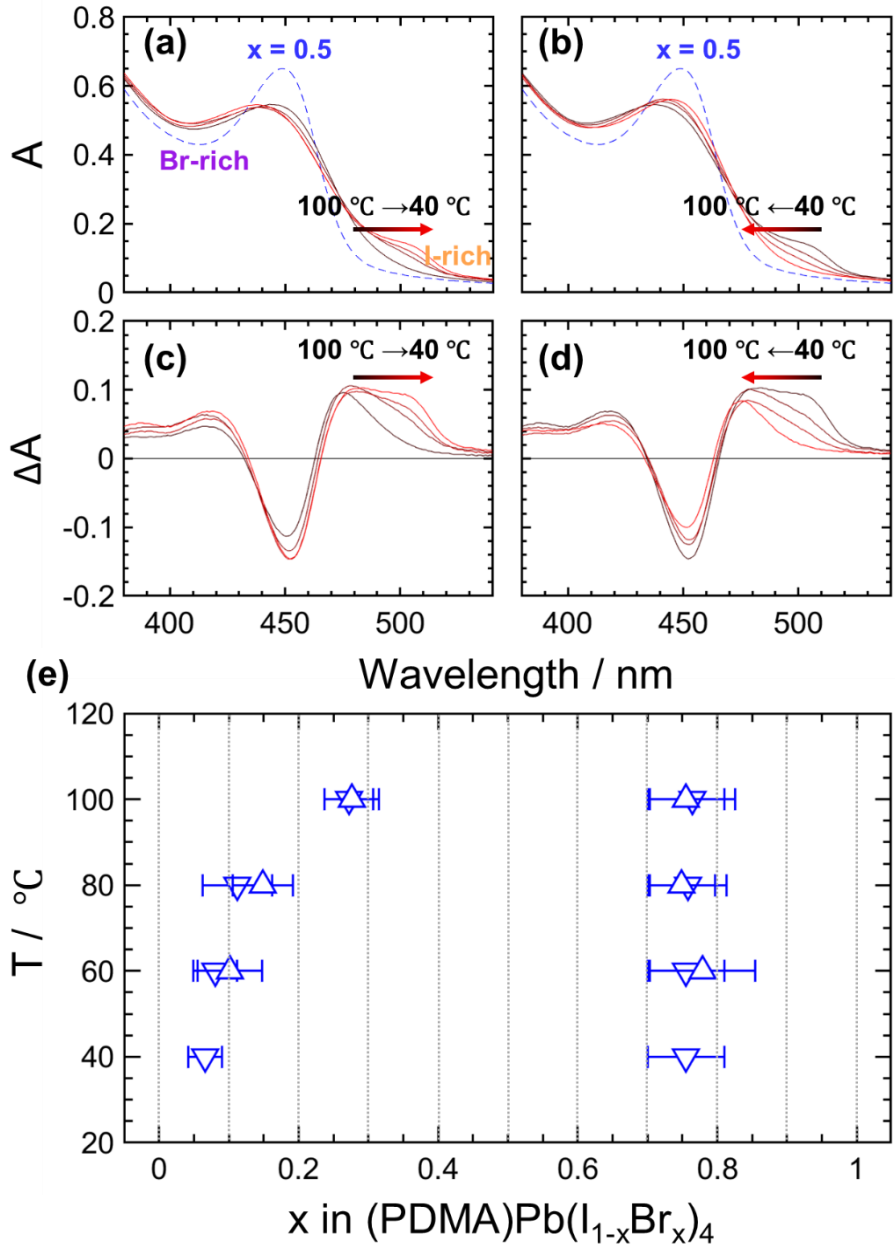

**Figure 3.** Temperature-dependent UV-Vis measurements performed on a (PDMA)Pb($I_{0.5}Br_{0.5}$)$_4$ thin film (with PMMA encapsulation) under light (1.5 mW/cm$^2$) at temperature varying as follows: (a) 100°C → 80°C → 60°C → 40 °C, (b) 40 °C → 60°C → 80 °C →100 °C (displayed spectra were taken after waiting at each temperature for 20 hours). The gradient color from black to red denotes the sequence of the temperature step. The dashed blue lines indicate the absorption spectrum of the pristine (PDMA)Pb($I_{0.5}Br_{0.5}$)$_4$ phase ($x_{initial}$ = 0.5) at 100°C. (c–d) Change in absorbance Δ$A$ spectra (obtained by subtracting from each absorption spectrum the spectrum of the pristine sample). (e) Photo-miscibility-gap extracted from temperature dependent UV-vis measurements shown in a–d; Downward-

pointing (▽) and upward-pointing (△) triangles denote data collected during a downward and upward temperature scan respectively (for details on the method to evaluate the photo-miscibility gap, see text and **Section S5** of the Supporting Information).

To provide a quantitative evaluation of the temperature dependence of the de-mixed compositions under light, we extract characteristic absorption wavelengths associated with the Br-rich and I-rich domains. For each temperature condition, we assign the peak position obtained from the $\Delta A$ spectra at short wavelengths and longtime scales to the signature for the Br-rich phase. For the I-rich phase, we consider the wavelength at which a (negative) peak in slope of $\Delta A$ vs wavelength is detected. For detailed information on the method used to extract composition information, see **Section S5** of the supporting information (**Figure S7-S11**). The data (summarized in **Figure 3e**) show that, by decreasing the temperature, the gap for the de-mixed compositions widens. The photo de-mixed composition for the I-rich domain approaches the value corresponding to the end member composition (x = 0). The feature associated to the Br-rich domain undergoes much more limited changes over the probed temperature range. This indicates that the degree of de-mixing, and therefore the range of immiscible compositions, increases in width with decreasing temperature, which is consistent with the decrease in the molar entropy of mixing contribution to the free energy when decreasing temperature. As discussed above, the analysis of the upward temperature scan (shown by △) displays an analogous trend. The critical temperature of this photo-miscibility gap, above which the entropic contribution would be large enough to prevent de-mixing, is an important parameter. We note that UV-vis measurements performed on a sample with $x_{initial}$ = 0.5 under light at 150 °C showed the appearance of absorption features associated with Br-rich and I-rich phases. While other irreversible absorption changes are also detected, this experiment indicates the occurrence of photo de-mixing (**Figure S18**) and suggests a critical temperature higher than 150 °C.

Regarding the temperature dependent UV-vis experiment shown in **Figure 3**, XRD measurements were also taken after each temperature step to evaluate the influence of light and temperature on the film's structural properties. Compared to the XRD measurement performed on the pristine sample, we observed a decrease of peak intensity and broadening of peak width upon illumination during the first temperature step. However, negligible changes in the XRD pattern were detected for all of the following temperature steps (**Figure S29**). This result suggests that, while the structural properties of the film do change upon exposure to light, a new stable structure is reached and maintained throughout the experiment. To further test that this initial change in properties reaches a stable condition, we ran an analogous experiment but conducted on different pristine samples for each of the temperature values considered above (see Supporting Information **Section S6**). These experiments showed comparable results (**Figure S12-S14**) and further validated the shape of the photo-miscibility gap. In addition, from the analysis of the temperature dependent kinetics for the change in the absorbance obtained during photo de-mixing (**Figure S15–S16**), we obtain an activation energy of 0.94 eV (**Figure S17**). This value is larger than the value reported for photo de-mixing in 3D mixed halide perovskite case.[11a, 27] The XRD measurements performed on these films show a decrease in intensity and broadening of some diffraction peaks after photo de-mixing (**Figure S30**), indicating a decrease in crystallinity, as also mentioned above. Interestingly, this phenomenon seems to affect predominantly peaks that correspond to the (h00) orientation, suggesting an anisotropic change in crystallinity. The effect was also more pronounced for samples that were illuminated at higher temperatures (**Figure S31**).

In order to check the sensitivity of the boundaries of the extracted photo-miscibility-gap to the initial composition of the 2D perovskite thin film, we have performed temperature-dependent photo de-mixing also on films with $x_{initial}$ = 0.4 and 0.6, analogously to the experiment shown in **Figure 3** (see **Figure S24**). We observe that the composition associated to the I-rich phase forming after photo de-mixing is rather insensitive to the value of $x_{initial}$. In contrast, the composition datasets extracted for the Br-rich phase seems to vary significantly depending on the pristine film composition. While the error associated with such estimates is relatively large for some of the data (see more details in **Section S5** of the Supporting Information), this may suggest slight variations in the steady-state photo de-mixed compositions depending on the starting condition.

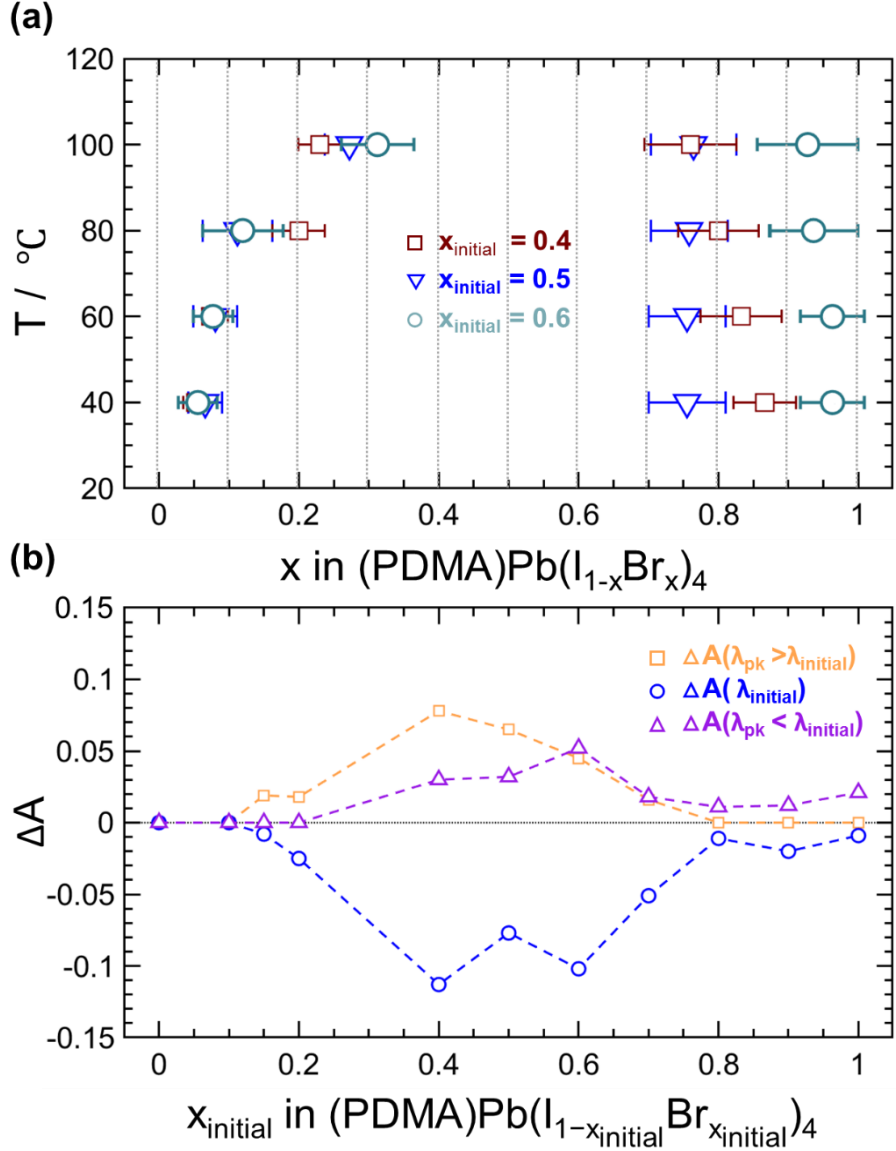

**Figure 4.** Photo de-mixing study on thin films of (PDMA)Pb($I_{1-x}Br_x$)$_4$ with different initial composition performed under illumination (1.5 mW cm$^{-2}$). (a) Photo-miscibility-gap extracted from temperature dependent UV-vis. Films with $x_{initial}$ = 0.4 (brown), $x_{initial}$ = 0.5 (blue) and $x_{initial}$ = 0.6 (cyan) were considered. The data correspond to measurements of the films taken after 20 hours at each temperature under continuous illumination. (b) Changes in absorbance $\Delta A$ measured at 100°C and after 2 hours of illumination plotted as function of the films' initial composition. The dataset indicated by the blue circles displays the bleach in the absorption at $\lambda_{initial}$ (peak wavelength of the pristine film). The yellow squares and the purple triangles indicate the peak values of $\Delta A$ features that emerged during illumination at wavelengths that are longer or shorter than $\lambda_{initial}$, respectively. When no feature was detected, a value of $\Delta A = 0$ is shown.

Based on this observation, we further test the accuracy of the methods described above, by investigating the phase stability of samples with $x_{initial}$ spanning the full range between 0 and 1. We consider the photo de-mixing of samples at 100°C after 2 hours under 1.5 mW cm$^{-2}$ illumination intensity. **Figure 4b** illustrates the changes in absorbance associated to the bleach of the main excitonic peak (at $\lambda = \lambda_{initial}$) as well as the peaks in $\Delta A$ of features emerging due to red-shift or blue-shift of the absorption spectrum. The data reveal that changes in absorbance, and therefore possible phase instability, under light occur for films with $x_{initial} \geq 0.15$ (see also **Figure S19–S23**). However the emergence of a clear Br-rich phase is only observed for $x_{initial} > 0.2$, which is consistent with our previous analysis at this temperature (see **Figure 3e** and **4a**). Furthermore, for samples with $x_{initial} \geq 0.8$, no feature associated to an I-rich phase compared to the initial composition was detected, suggesting that no de-mixing occurred in such cases. This boundary falls within the uncertainty region of the data in **Figure 4a**. We note that a slight blue-shift in absorption still occurs for $x_{initial} \geq 0.8$ and even for $x_{initial} = 1$.

While the methods described above have the potential to map the miscibility gap of samples under a given illumination intensity, we acknowledge that the situation is more complicated compared to a study of a system in the dark at equilibrium. De-mixing is likely to depend on the concentration of photo-generated electronic charge carriers, and this may vary during the experiment, for example due to changes in recombination. To test the sensitivity of the measured photo-miscibility-gap to the bias light, we ran a similar experiment as the one shown in **Figure 3** with 10-times higher light intensity. We observe slight differences in the shape of the temperature-dependent absorbance of the thin film under the two light conditions, with a more pronounced excitonic signature associated to the I-rich domain at low temperatures. However, the analysis (**Section S8** in the Supporting Information, **Figure S25-S26**) using the method described above results in surprisingly similar de-mixed compositions for the two different light intensities, suggesting limited variation in the stable compositions while varying the optical perturbation in this range.

Finally, we investigated the effect of encapsulation of the sample on the photo de-mixing behavior of these 2D perovskite thin films (**Section S9** in the Supporting Information, **Figure S27**). Similar to the previous reports on 3D perovskite films,[28] encapsulation was found to be essential to suppress the degradation of the I-rich phase, improving the reversibility of photo de-mixing and dark re-mixing. Interestingly, the non-encapsulated sample demonstrated a widening of the photo-miscibility gap when compared to a reference sample encapsulated with PMMA (data in **Figure 3**). Changes in absorbance for the non-encapsulated sample highlight a drop in the signature associated to I-rich domains, suggesting the presence of a loss mechanism for such phase, likely due to $I_2$ excorporation. These observations highlight that further understanding of the role of stoichiometry on the thermodynamics of photo de-mixing remains a key question and point to the importance of encapsulation in the analysis of the behavior of halide perovskite materials.

Based on our analysis, we have shown the tunability of the optical properties of the 2D Dion-Jacobson perovskite films through halide substitution. Furthermore, our results on photo de-mixing of these compounds demonstrate that light-induced ion transport effects play an important role in confined 2D Dion-Jacobson systems, similarly to the case of 3D mixed-halide perovskites and 2D Ruddlesen-Popper. Our data also indicate that no miscibility gap is present in the dark for these systems and that a thermodynamic rather than kinetic origin underlies the process of photo-de-mixing in 2D films, similarly to their 3D counterparts. Our careful study of the photo-miscibility gap establishes the phase thermodynamics of 2D mixed-halide perovskites out-of-equilibrium experimentally. The origin of the phase instability under light may be related to selective self-trapping, as suggested for 3D halide perovskites.[16] According to such model, the energies related to defect interaction with electronic charge carriers, and specifically the difference in hole trapping behavior for iodide and bromide rich phases, would be key to the driving force for de-mixing. However, unlike 3D systems, the large exciton binding energy in 2D perovskites[29] investigated here may also have influence on the interaction between the electronic charge carriers and ionic defects. In addition, the peculiar shape of the photo-miscibility-gap on the Br-rich side may be related to a more complex mechanism during de-mixing, such as ordering effects in the Br-rich phase. Such a "steep" shape of the miscibility gap has also been observed in calculations of 3D mixed iodide and bromide systems of $CsPbI_{1-x}Br_x$ for the I-rich phase for x = 1/3.[30]

**Conclusion**

In summary, we demonstrated that 2D Dion-Jacobson mixed-halide perovskites undergo photo de-mixing with direct conversion from the pristine $(PDMA)Pb(I_{0.5}Br_{0.5})_4$ to I-rich and Br-rich phases under illumination. In the dark, the photo de-mixed phases re-mix with almost complete reversibility of their optical properties, indicating full miscibility in the dark. Interestingly, an irreversible loss in crystallinity also takes place upon illumination. The temperature-dependence of the photo de-mixed phases evidenced a photo-miscibility-gap in a temperature range relevant to applications of these materials in optoelectronic devices. This work thereby provides a practical approach with multiple methods for defining the thermodynamically stable compositions of hybrid perovskite mixtures under light, presenting the basis for developing a fundamental understanding of photo de-mixing in 2D mixed-halide perovskites. These findings will aid compositional engineering related to halide mixtures to enable optimization of optoelectronic devices as well as the development of other emerging systems exploiting opto-ionic effects.

**Acknowledgements**

We are grateful to Helga Hoier for XRD measurements, Florian Kaiser and Rotraut Merkle for technical assistance. DM is grateful to the Alexander von Humboldt Foundation for funding. This work was performed within the framework of the Max Planck-EPFL Center for Molecular Nano-science and Technology.

Supporting Information

# Photo de-mixing in Dion-Jacobson two-dimensional mixed halide perovskites

Ya-Ru Wang,[1] Alessandro Senocrate,[1] Marko Mladenović,[2] Algirdas Dučinskas,[1,3] Gee Yeong Kim,[1] Ursula Röthlisberger,[2] Jovana V. Milić,[3] Davide Moia,[1]* Michael Grätzel,[1,3] Joachim Maier[1]*

[1] Max Planck Institute for Solid State Research, Stuttgart, Germany.

[2] Laboratory of Computational Chemistry and Biochemistry, Institute of Chemical Sciences and Engineering, École Polytechnique Fédérale de Lausanne (EPFL), Lausanne, Switzerland.

[3] Laboratory of Photonics and Interfaces, Ecole polytechnique Fédérale de Lausanne (EPFL), Lausanne, Switzerland.

*d.moia@fkf.mpg.de; office-maier@fkf.mpg.de

## Table of Contents

**S1 Materials and methods**

*Materials:* lead iodide (PbI$_2$) and lead bromide (PbBr$_2$) were purchased from Alfa Aesar. 1,4-phenylenedimethanammonium iodide ((PDMA)I$_2$) spacer, 1,4-phenylenedimethanammonium bromide (PDMA)Br$_2$) spacer were synthesized according to the reported procedure[1] and the one described below. Dimethylsulfoxide (DMSO), Dimethylformamide (DMF) and Poly(methyl methacrylate (PMMA) were purchased from Sigma-Aldrich.

*Synthesis of (PDMA)Br$_2$:* The solution of 1,4-phenylenedimethanamine (1.04 g, 7.64 mmol) was suspended in EtOH (10 mL) at ambient temperature and treated with 3.0 mL HBr (48%, 3.5 equiv). The reaction mixture was stirred over 3 h at ambient temperature and dried under vacuum. The mixture was re-dispersed into diethyl ether (100 mL), filtered, extensively washed with from the mixture of diethyl ether and isopropanol, and dried under vacuum to afford (PDMA)Br$_2$ (2.2 g, 97%) as a white powder.

$^1$H NMR (400 MHz, (CD$_3$)$_2$SO): $\delta$ = 8.28 (s, 6H), 7.51 (s, 4H), 4.06 (d, *J* = 5.6 Hz, 4H) ppm; $^{13}$C NMR (101 MHz, (CD$_3$)$_2$SO): $\delta$ = 134.61, 129.56, 42.29 ppm; HRMS (ESI+/QTOF): *m/z* (%) 120.0812 (100, [*M*]$^+$, calcd. for C$_8$H$_{10}$N$^+$: 120.0808); 78.9179 (100, [M]$^-$, calcd. for Br$^-$: 78.9189). (See NMR spectra in **Figure S2** and **S3**).

*Synthesis of (PDMA)Pb(I$_{1-x}$Br$_x$)$_4$ precursor:* the precursor solutions with Br content of 0%, 10%, 20%, 30%, 40%, 50% 60%, 70%, 80%, 90% and 100% were prepared following the relative stoichiometry of the halides. Specifically, 0.33 M (PDMA)PbI$_4$ or (PDMA)PbBr$_4$ solutions were prepared by dissolving 0.33 mmol (PDMA)I$_2$ / (PDMA)Br$_2$ and 0.33 mmol PbI$_2$ (PbBr$_2$) in the solvent mixture of DMF and DMSO (3:2, (v:v), 200 μL). The I-rich mixed binary anion lead halide perovskites precursors (x = 0.1, 0.2, 0.3, 0.4, 0.5) were prepared by dissolving (PDMA)I$_2$, PbI$_2$, and PbBr$_2$ with stoichiometry of 1: (1-2x): 2x. Similarly, the Br-rich mixed binary anion lead halide perovskites precursors (x = 0.6, 0.7, 0.8, 0.9) were prepared by dissolving (PDMA)Br$_2$, PbI$_2$, and PbBr$_2$ with stoichiometry of 1: (2-2x): (2x-1).

***Preparation of (PDMA)Pb(I$_{1-x}$Br$_x$)$_4$ films:*** the film preparation procedure was conducted in Ar-filled glovebox with controlled atmosphere (O$_2$ and H$_2$O < 0.1 ppm). The (PDMA)Pb(I$_{1-x}$Br$_x$)$_4$ films were deposited on quartz substrate by spin coating the precursor solution at 9 rps and 66 rps for 2s and 48s. The as deposited films were annealed at 150 °C for 10 minutes. A PMMA encapsulation layer was deposited on perovskite films for encapsulation experiments by drop casting PMMA solution (10 mg/ml, dissolved in chlorobenzene). The coated sample was dried at 40 °C for 2 hours in glovebox.

***Ultraviolet–visible spectroscopy experiment under controlled conditions (UV-Vis):*** UV - Vis experiments were conducted using a Shimadzu UV-2600 spectrometer. The setup was modified to achieve good control of temperature and atmosphere. An O$_2$ sensor was used in this experiment to monitor the O$_2$ concentration in the exhaust gas line during the measurements (< 30 ppm O$_2$ was maintained for all the measurements). Heating pads were used to control the temperature in the system and a thermocouple was used to monitor the temperature of the sample. A LED lamp (MCWHF$_2$, Thorlab) coupled to an optical fiber is used for the illumination experiments. Light intensity was measured directly at the exit of the optical fiber using a thermal power meter (PM100 D, Thorlab). The modulation of the bias light to allow for illumination and UV-Vis measurements is shown in **Figure S1**. For each cycle, light was switched on for 900 s followed by a 300 s dark window for UV-Vis measurements to be taken.

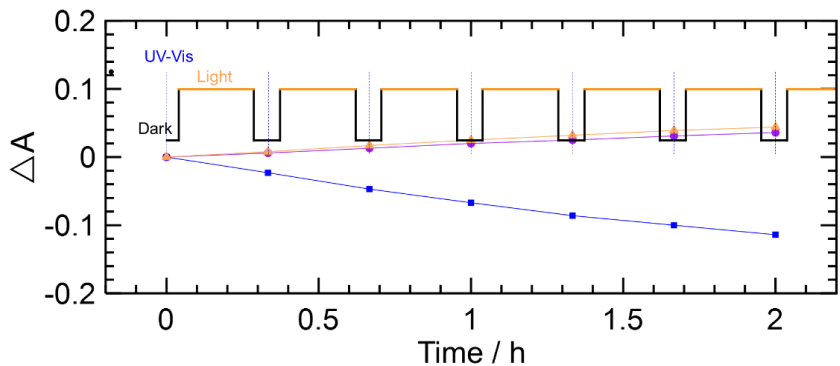

**Figure S1.** Representation of the bias light modulation during photo de-mixing experiments. The blue dashed lines indicate the times where the UV-Vis measurements were taken. The blue,

yellow and purple data points correspond to the changes in absorbance evaluated by means of the UV-vis measurements.

*X-Ray Diffraction (XRD):* All the XRD patterns were acquired by a PANalytical diffractometer of Empyrean Series 2 (Cu Kα radiation, 40 kV, 40 mA) equipped with a parallel beam mirror and a PIXcel 3D detector. All spectra were recorded in normal scan (ω=2°) configuration. The samples were mounted in a polycarbonate domed sample holder for protection from ambient atmosphere during measurements. For temperature dependence experiments the samples were mounted in a high temperature oven chamber (HTK 1200N) from Anton Paar equipped with a EUROTHERM 2604 controller.

*Ab initio calculations* based on the Generalized Gradient Approximation (GGA) of Density Functional Theory (DFT) (PDMA)PbI$_4$, (PDMA)Pb(I$_{0.5}$Br$_{0.5}$)$_4$, and (PDMA)PbBr$_4$ were performed using the Quantum Espresso code.[2] The Perdew–Burke–Ernzerhof functional revised for solids (PBEsol)[3] was chosen based on the benchmarks performed on the (PDMA)PbI$_4$ system, in our previous study.[4] To calculate band gaps, we employed a higher level of approximation using the hybrid functional PBE0[5] together with explicit inclusion of spin-orbit coupling (SOC) effects. To model valence-core electron interactions, ultrasoft pseudopotentials were employed with a plane wave basis set with kinetic energy cutoffs for the wavefunction and for the density of 50 Ry and 350 Ry, respectively. For band gap calculations, norm-conserving pseudopotentials with 80 Ry wavefunction cutoff and 320 Ry density cutoff were employed. The Brillouin zone was sampled by a 3×3×3 k-points grid. For band gap calculations, norm-conserving pseudopotentials with 80 Ry wavefunction cutoff and 320 Ry density cutoff were employed. Free energies were calculated by employing the PBE functional[6] together with D3 dispersion corrections.[7]

*NMR Spectra:* $^1$H NMR Spectra were recorded on a Bruker DRX 400 instrument operating at 400 MHz at 298 K. $^{13}$C NMR spectra were recorded on a Bruker DRX 400 operating at 100 MHz. Multiplicities are reported as follows: bs (broad singlet), s (singlet), d (doublet), and m (multiplet).   Chemical shifts δ (ppm) were referenced to the internal solvent signals.

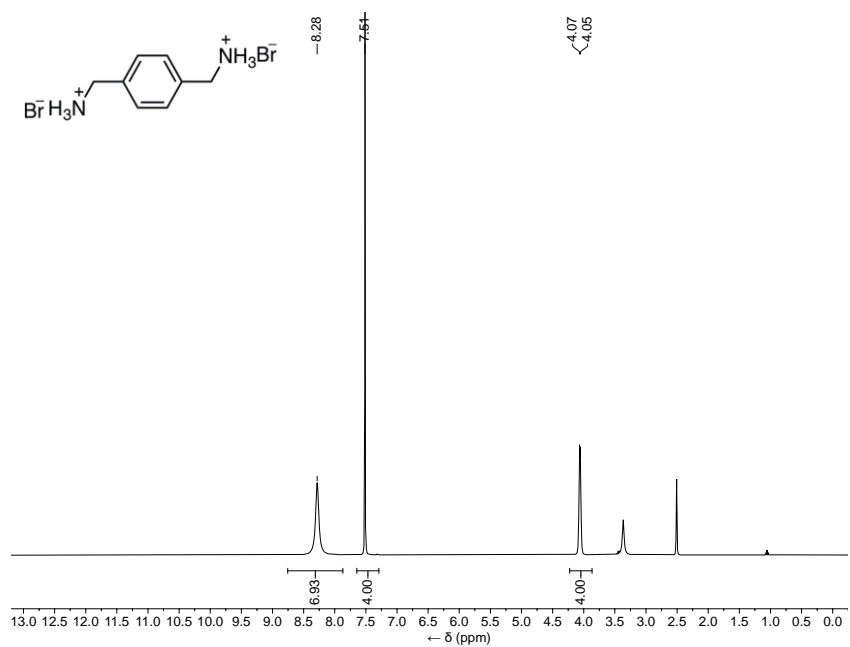

**Figure S2.**   ¹H NMR of **(PDMA)Br₂** in (CD₃)₂SO (400 MHz, 298 K)

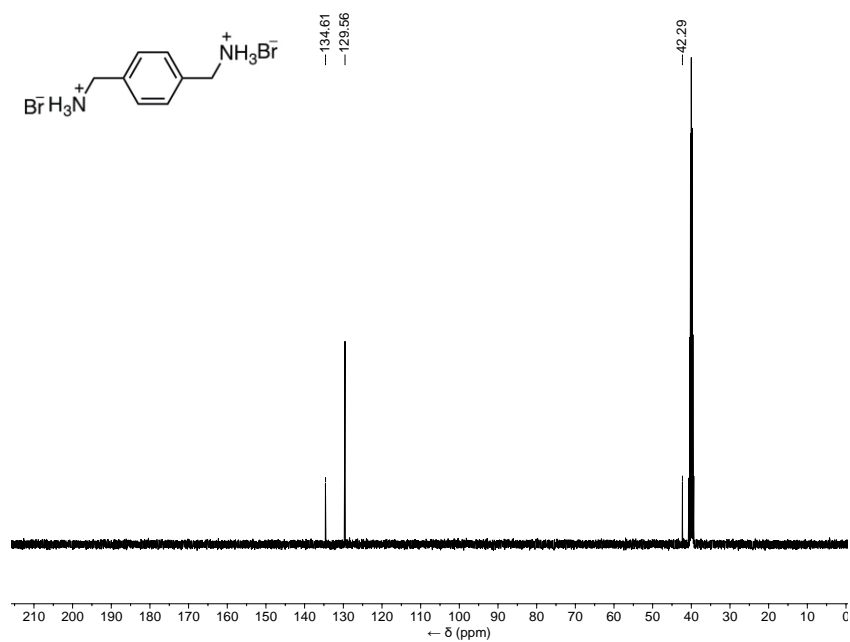

**Figure S3.**   ¹³C NMR of **(PDMA)Br₂** in (CD₃)₂SO (101 MHz, 298 K)

## S2. Calculations

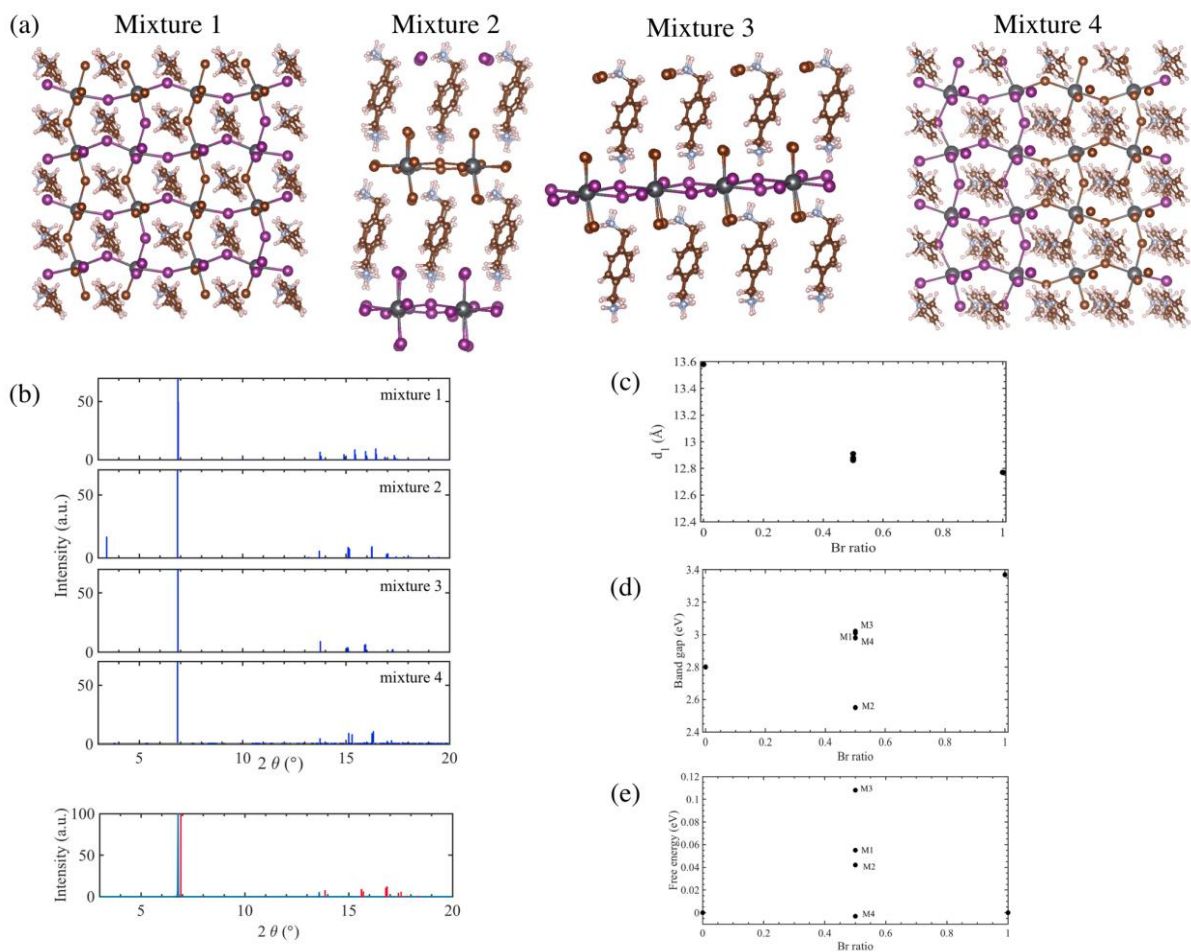

**Figure S4 (a)** Atomic structures of (PDMA)Pb(I$_{0.5}$Br$_{0.5}$)$_4$ with different mixing patterns. (b) XRD patterns of the DFT-**optimzed** structures of (PDMA)Pb(I$_{0.5}$Br$_{0.5}$)$_4$ (top) and a mixture of (PDMA)PbI$_4$ and (PDMA)PbBr$_4$. (bottom). (c) Spacing between inorganic layers (d$_1$) for (PDMA)PbI$_4$, (PDMA)Pb(I$_{0.5}$Br$_{0.5}$)$_4$, and (PDMA)PbBr$_4$. (d) PBE0+SOC band gaps of (PDMA)PbI$_4$, (PDMA)Pb(I$_{0.5}$Br$_{0.5}$)$_4$, and (PDMA)PbBr$_4$. (e) Free energy of mixing for different (PDMA)Pb(I$_{0.5}$Br$_{0.5}$)$_4$ structures, calculated as $\Delta F = \Delta H - T\Delta S$, where $\Delta H$ is the internal energy difference of the mixtures and the pure compounds and $T\Delta S = -k_B T(x \ln x + (1-x)\ln(1-x))$ is the entropic contribution ($T = 350K$ and $x = 0.5$).

To gain more insights on the atomic structure of (PDMA)Pb(I$_{0.5}$Br$_{0.5}$)$_4$, we investigated structures with different mixing patterns of I and Br ions. As shown in **Figure S4a,** mixture 1 represents a homogeneous mixture of I and Br ions. In the case of mixture 2, inorganic layers fully consist of either I or Br ions, which alternate in the direction perpendicular to the layers. On the other hand, in mixture 3, Br ions reside on dangling sites, while I ions reside in the central part of the inorganic layer. Finally, mixture 4 represents an inhomogeneous mixing of I and Br ions within the same inorganic layer where I- and Br-rich regions are fully separated. The XRD patterns of all investigated models exhibit strong low-angle peaks around 6.9° with no significant differences between the structures (**Figure S4 b**), with the exception of the occurrence of a low-angle peak at around 3.4° for mixture 2, which is due to the periodic alternation of I- and Br-rich inorganic layers. Small differences in XRD patterns are consistent with similar values of interlayer spacing ($d_1$) for all models of $(PDMA)Pb(I_{0.5}Br_{0.5})_4$ **(Figure S4 c)**, which are slightly above the measured value of 12.34 Å. Band gaps of the investigated models of $(PDMA)Pb(I_{0.5}Br_{0.5})_4$, shown in **Figure S4 d**, mainly lie between the values for pure compounds, in a good agreement with the experimental measurements, that give a band gap of around 2.75 eV. A notable exception from this trend is the band gap of mixture 2 that is significantly lower than the band gaps of pure compounds, indicating that such a structure is not consistent with the measured structure of $(PDMA)Pb(I_{0.5}Br_{0.5})_4$. Finally, we calculated free energies of mixing for the $(PDMA)Pb(I_{0.5}Br_{0.5})_4$ systems with respect to the pure compounds (**Figure S4 e**). The positive free energy values for most of the mixtures suggest their enthalpic instability, which may serve as a possible explanation for the observed de-mixing upon exposure to light. However, the inhomogeneous mixture (mixture 4) exhibits the lowest and negative free energy, indicating the possibility of stabilizing a mixed phase at higher temperatures.

## S3. 2D Mixed halide perovskites (PDMA)Pb(I$_{1-x}$Br$_x$)$_4$

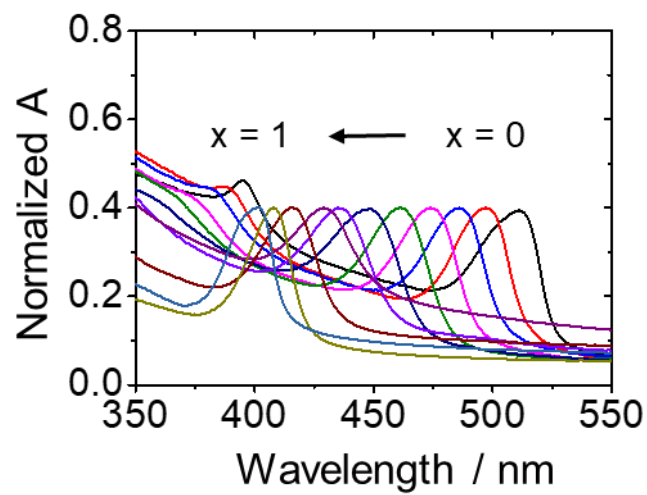

**Figure S5.** Normalized absorbance spectra for (PDMA)Pb(I$_{1-x}$Br$_x$)$_4$ thin films with x = 0, 0.1, 0.2, 0.3, 0.4, 0.5, 0.6, 0.7, 0.8, 0.9, 1.

## S4. UV-Vis evolution under light for (PDMA)PbI$_4$ and (PDMA)PbBr$_4$

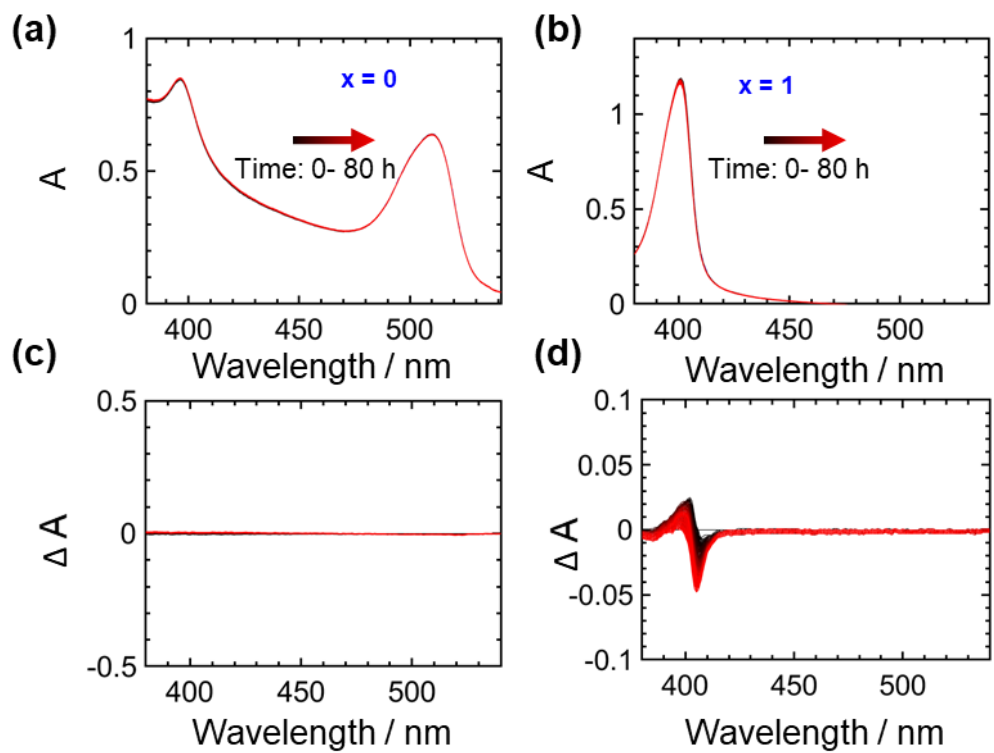

**Figure S6.** UV-vis spectra for (PDMA)PbI$_4$ (a) and (PDMA)PbBr$_4$ (b) thin films at 80 °C (with PMMA encapsulation in Ar atmosphere) with 1.5 mW/cm$^2$ for 80 h. No absorption change is detected for (PDMA)PbI$_4$. For (PDMA)PbBr$_4$, a slight blue shift in absorption is detected. (c) and (d) are the corresponding ΔA spectra (obtained by subtracting the reference spectrum of the pristine sample from each absorption spectrum).

## S5. Methods for identification of the Br-rich and I-rich phases

Here, we discuss the methods used for extracting the characteristic absorption wavelengths for Br-rich and I-rich phases referred to the photo de-mixing experiments in the main text. The estimate of the compositions associated to these features is also discussed in detail below.

*Estimation of the I-rich phase composition*

**Figure S7a, b** show the absorbance and the change in absorbance evolution of a 2D DJ perovskite with x = 0.5 performed at 100 °C for 20 hours. Both graphs highlight the emergence of features that can be assigned to I-rich and Br-rich phases as discussed in the main text. In our analysis, we assume that the feature due to the I-rich phase is only marginally influenced by the changes in absorbance due to the disappearance of the pristine phase (x = 0.5). A clear signature that can be associated to an excitonic peak of the I-rich phase is only occasionally visible. Therefore, we extract a characteristic wavelength for the I-rich absorption signature defined based on the (negative) peak of the slope of the ΔA signal with respect to wavelength (**Figure S7c**). The calibration used to determine the composition of the I-rich phase will be discussed below.

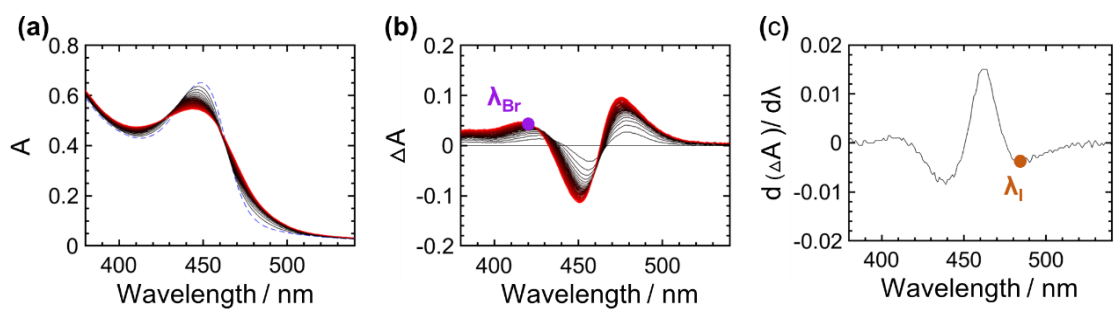

**Figure S7.** (a) Raw UV-vis (b) ΔA and (c) $d\Delta A/d\lambda$ spectra for photo de-mixing performed at 100°C. The graphs show examples of the characteristic wavelengths corresponding to Br-rich ($\lambda_{Br}$) and I-rich ($\lambda_I$). $\lambda_{Br}$ was obtained from the peak of the ΔA at short wavelengths, while $\lambda_I$ was obtained from the peak of the slope of ΔA.

**Figure S8** displays the derivative of the change in absorbance for the sample discussed in Figure 3 in the main text, after 20 hours of illumination and for the four temperature tested in the experiment. The derivative of ΔA does not always show a sharp peak, also due to the noise in the data after differentiation. We determine the peak position by smoothing the data using the Savitzky-Golay method (15 point window). We then evaluate the error associated to the determination of such peak position considering the root mean square (rms) of the residual obtained from the smoothing and evaluate the wavelength range where the value of the smoothed data is close to the peak value within 2 times the rms.

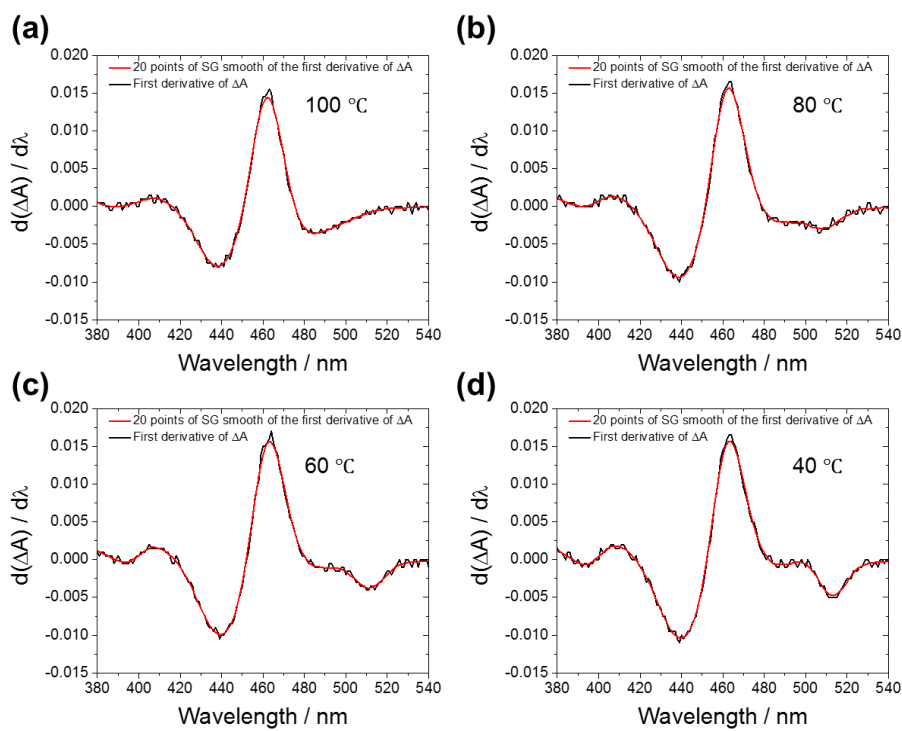

**Figure S8.** Derivative of ΔA with respect to wavelength for the data shown in Figure 3 in the main text collected at different temperatures: (a) 100 °C; (b) 80 °C; (c) 60 °C; (d) 40 °C. The black and the red lines denote the raw data and the smoothed data. The latter were obtained using the Savitzky-Golay method (number of points used for smoothing window = 15).

Next, in order to correlate the characteristic wavelengths of the I-rich phases with their corresponding compositions, UV-Vis measurements are conducted for reference samples of (PDMA)Pb(I$_{1-x}$Br$_x$)$_4$ thin films ($x$ = 0, 0.1, 0.2, 0.3, 0.4, 0.5, 0.6, 0.7, 0.8, 0.9, 1) (see **Figure S5**). Below, we use A$_x$ to refer to the spectrum of a film fabricated with a composition expressed by the value of $x$. From the negative peak in the derivative of A$_x$ with respect to wavelength observed at long wavelength we are able to determine a calibration function to extract an estimate of the I-rich composition obtained after de-mixing. In order to check that changes in absorption due to the reduction in the absorption of the pristine composition x = 0.5 during the photo de-mixing experiment does not affect the calibration, we also check the position of the peak in the derivative of $\Delta A_x$ defined as $\Delta A_x = A_x - A_{0.5}$ (see data in **Figure S9a**). We note that such value is the same as the (negative) peak of the slope of $A_x$, validating the approach used in the main text to extract $x$ for the I-rich phase from the measured $\Delta A$ during photo de-mixing. **Figure S9b** shows the corresponding calibration data to which a line was fitted and used to determine the composition of the I-rich phase in the main text and in this document.

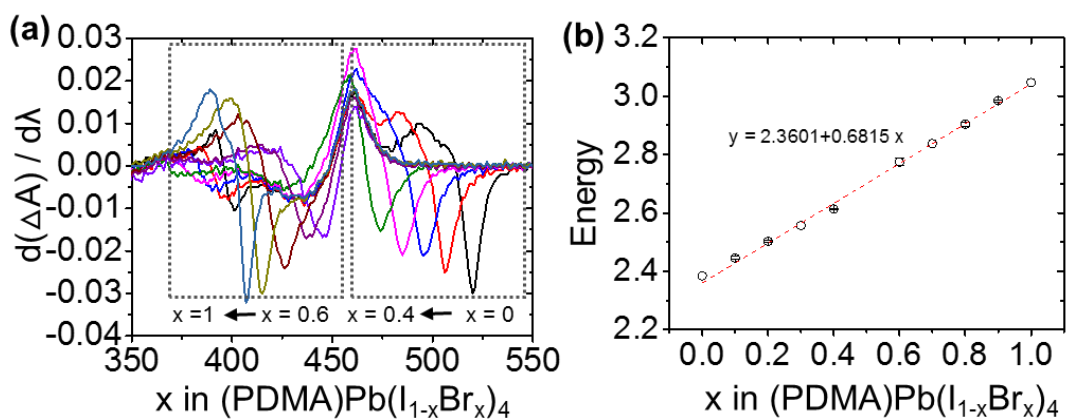

**Figure S9.** (a) Derivative of $\Delta A_x$ spectra for $(PDMA)Pb(I_{1-x}Br_x)_4$ thin films with x = 0, 0.1, 0.2, 0.3, 0.4, 0.5, 0.6, 0.7, 0.8, 0.9, 1 (shown in **Figure S5**). (obtained by subtracting $A_{0.5}$ from each spectrum). (b) Absorption energy as a function of bromide content corresponding to the absorption wavelengths obtained from the (negative) peak of the slope of $\Delta A_x$ shown in (a). The error bars are very small for this case and were determined with the same method as mentioned above for de-mixed spectra at different temperature.

*Estimation of the Br-rich phase composition*

The characteristic absorption wavelengths of Br-rich phases were determined based on the peak position of the $\Delta A$ spectra at short wavelengths (**Figure S7b**). We note that the absorption of the pristine compositions used in the experiments shown in the main text ($x_{initial}$ = 0.5, 0.4 and 0.6) is relatively flat in the relevant wavelength range, suggesting limited change in the shape and peak position related to the absorbance of the Br-rich phase upon variations in the absorption magnitude associated to the starting phase during the experiment. Nevertheless, we try to account for this effect and consider the influence from the emergence of the I-rich phase and from the consumption of the pristine phase during de-mixing. We calculated the expected $\Delta A$ spectrum based on the reference spectra shown in **Figure S5**. We considered different reference de-mixing scenarios involving all relevant combinations of I-rich and Br-rich contributions to $\Delta A$. As shown in Equation (1), we assume that, upon complete de-mixing, all of the pristine phase related to $x_{initial} = 0.5$ de-mixes and forms two phases with different concentrations and compositions, satisfying mass conservation. Then $\Delta A_{calc}$ is a result of the change in absorbance for the newly formed phases and the consumed pristine phase.

$$(PDMA)Pb(I_{0.5}Br_{0.5})_4 \rightarrow c_I(PDMA)Pb(I_{1-x_I}Br_{x_I})_4 + c_{Br}(PDMA)Pb(I_{1-x_{Br}}Br_{x_{Br}})_4 \quad \text{Eq. 1}$$

$$\Delta A_{calc} = c_I A_I + c_{Br} A_{Br} - A_{0.5} \quad \text{Eq. 2}$$

In Eq. 2, $c_I$ and $c_{Br}$ are molar ratios of the I- and Br-rich phases, and $A_I$ and $A_{Br}$ are the reference absorbance of the I- and Br-rich phases. By solving the equation of the mass balance shown in equation 1, the values of $c_I$ and $c_{Br}$ as function of the input compositions can be expressed as $c_I = \frac{1-2x_{Br}}{2x_I-2x_{Br}}$, $c_{Br} = \frac{2x_I-1}{2x_I-2x_{Br}}$, where $x_I$ and $x_{Br}$ indicate the compositions of the phases used in the calculation.

Since the temperature dependence of the de-mixed compositions yielded values of $x$ in the range 0–0.4 and 0.7–0.9 for I-rich phase and Br-rich respectively, here we investigated the calculated spectra based on the combination of the composition for I-rich (x = 0, 0.1, 0.2, 0.3, 0.4), Br-rich (x = 0.7, 0.8, 0.9). (**Figure S9**).

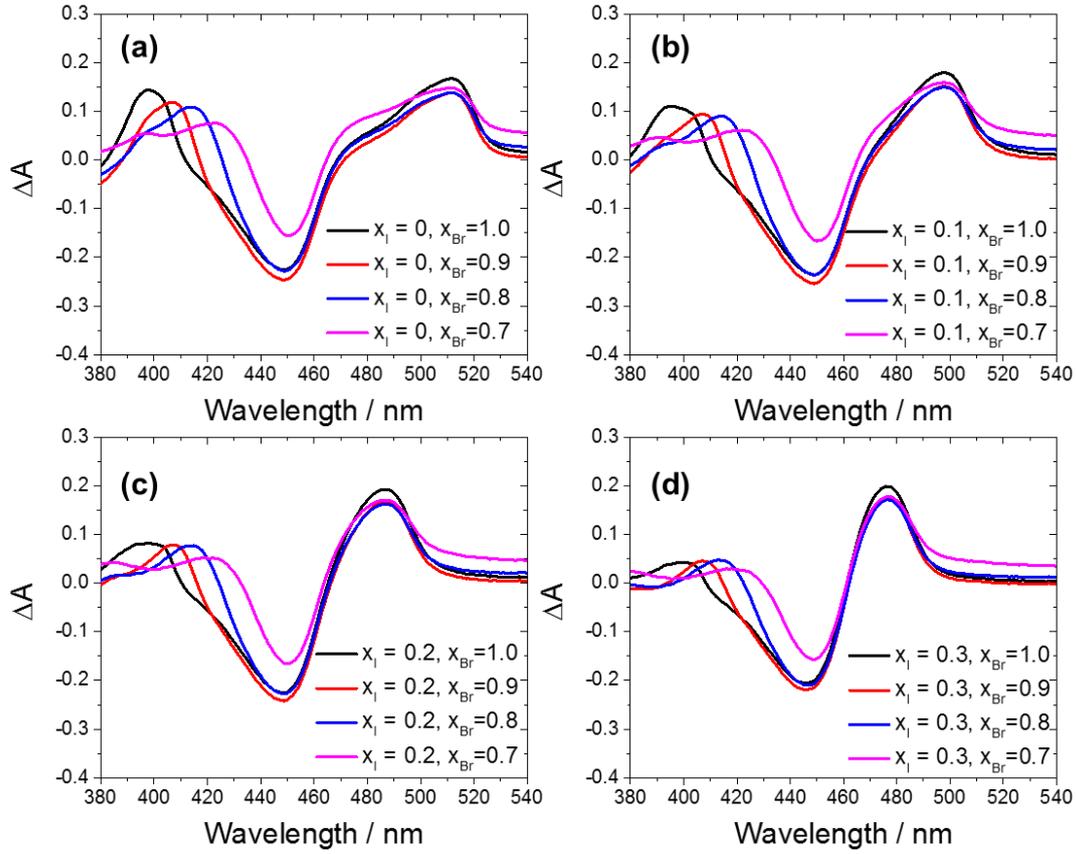

**Figure S10.** (a) $\Delta A_{calc}$ spectra assuming de-mixed composition of I-rich (x = 0, 0.1, 0.2, 0.3) and Br-rich (x=0.7, 0.8, 0.9, 1).

In all cases shown in **Figure S10**, we find that varying the value of $x_{Br}$ results in a clear trend in the peak position of $\Delta A_{calc}$ at short wavelengths. We note that the sharpness of the peak of the $\Delta A_{calc}$ at short wavelengths varies when considering different combinations of I-rich and Br-rich compositions. In order to account for this effect, we use the 95% of the peak intensity to determine the error associated to the peak position. In **Figure S11a** we compare the trends displayed in the four panels of **Figure S10** and show that the formation of the I-rich phase with different composition and the consumption of the $x_{initial} = 0.5$ phase have limited influence on the determination of the Br-rich composition in the calculated spectra. Based on this, we can apply this method to the experimental data. By using the calibration function shown in **Figure S11a**, the characteristic wavelengths of the Br-rich phase after de-mixing can be extracted and correlated with the phase composition. This allows us to map the right part of the photo-miscibility gap in **Figure 3.** The same method is used for determining the Br-rich compositions when using different initial composition of the pristine film (see reactions and mass conservation equations below).

$$(PDMA)Pb(I_{0.6}Br_{0.4})_4 \rightarrow c_I(PDMA)Pb(I_{1-x_I}Br_{x_I})_4 + c_{Br}(PDMA)Pb(I_{1-x_{Br}}Br_{x_{Br}})_4 \quad \text{Eq. 3}$$

$$\Delta A_{calc} = c_I A_I + c_{Br} A_{Br} - A_{0.4} \quad \text{Eq. 4}$$

$$c_I = \frac{x_{Br} - 0.4}{x_{Br} - x_I}, \quad c_{Br} = \frac{0.4 - x_I}{x_{Br} - x_I},$$

$$(PDMA)Pb(I_{0.4}Br_{0.6})_4 \rightarrow c_I(PDMA)Pb(I_{1-x_I}Br_{x_I})_4 + c_{Br}(PDMA)Pb(I_{1-x_{Br}}Br_{x_{Br}})_4 \quad \text{Eq. 5}$$

$$\Delta A_{calc} = c_I A_I + c_{Br} A_{Br} - A_{0.6} \quad \text{Eq. 6}$$

$$c_I = \frac{x_{Br} - 0.6}{x_{Br} - x_I}, \quad c_{Br} = \frac{0.6 - x_I}{x_{Br} - x_I}$$

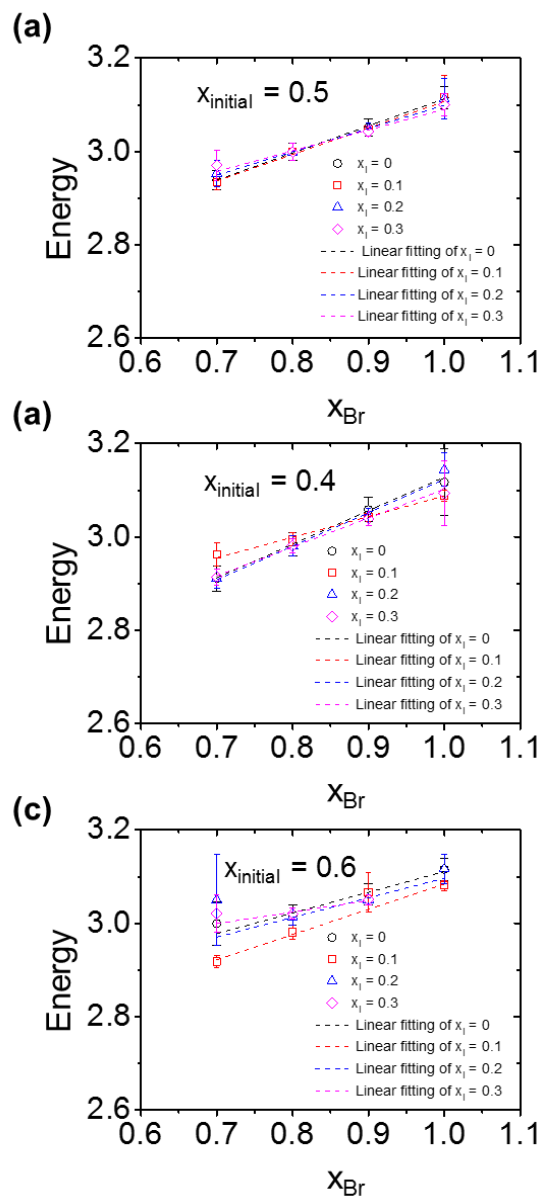

**Figure S11.** Calculated photon energy corresponding to the absorption wavelength associated to the Br-rich phase extracted from the peak of $\Delta A_{calc}$ for different combinations of I-rich and Br-rich compositions. The graphs refer to different values of the initial composition of the film $x_{initial}$. (a) $x_{initial} = 0.5$ (see also **Figure S10**), (b) $x_{initial} = 0.4$ and (c) $x_{initial} = 0.6$. The error bar of each point is determined by considering the wavelength range corresponding to 95% of the $\Delta A_{calc}$ peak intensity.

## S6. Reproducibility study of the photo-miscibility-gap

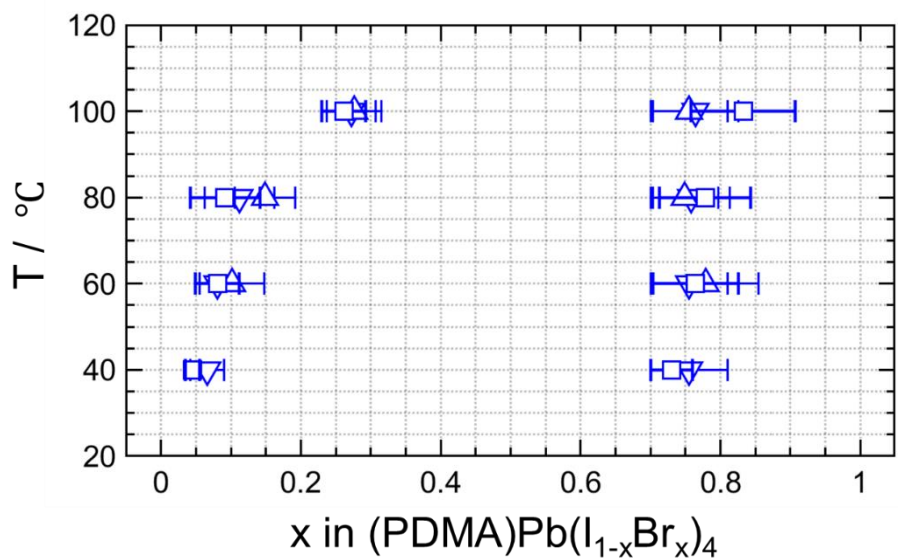

**Figure S12.** Photo-miscibility-gap mapped using multiple samples (squares, □) compared with the data shown in **Figure 3** in the main text obtained for a single sample (empty upwards △ and downwards pointing triangles ▽).

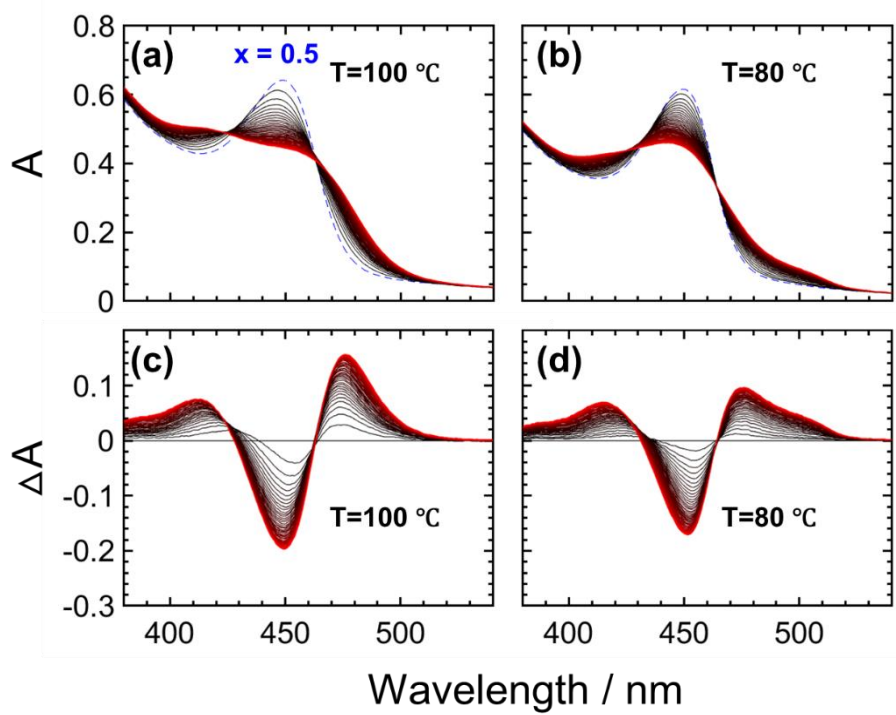

**Figure S13.** Temperature dependent UV-Vis spectra of a (PDMA)Pb(I$_{0.5}$Br$_{0.5}$)$_4$ thin film (with PMMA encapsulation) under light (1.5 mW/cm$^2$) (a) at 100 °C and (b) at 80 °C for 20 hours. (c) and (d) are the ΔA spectra (obtained by subtracting the reference spectrum of the pristine sample from each absorption spectrum).

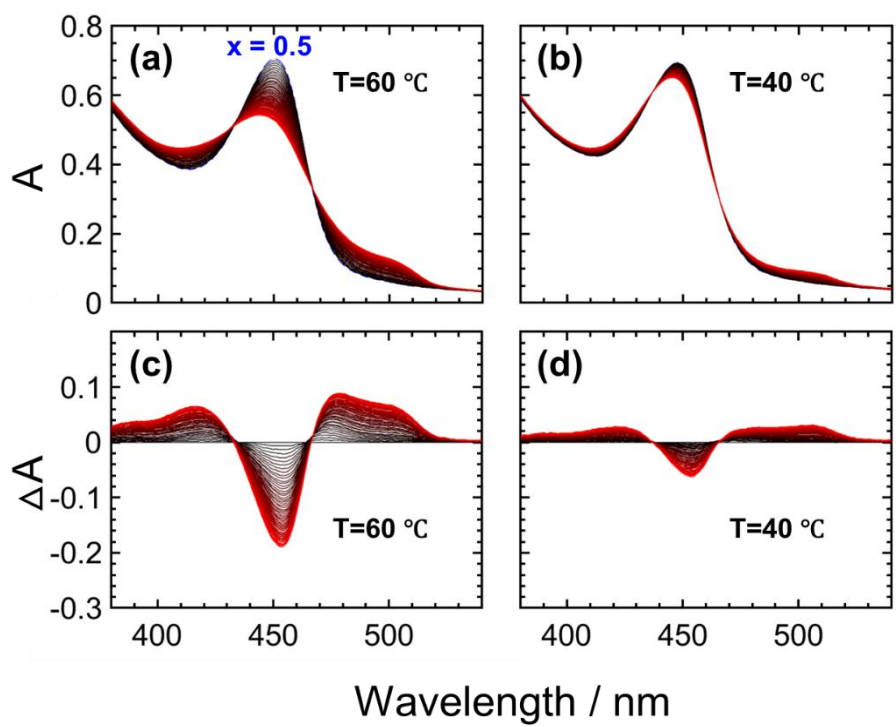

**Figure S14.** Temperature dependent UV-Vis spectra of (PDMA)Pb(I$_{0.5}$Br$_{0.5}$)$_4$ thin film (with PMMA encapsulation) under light (1.5 mW/cm$^2$) (a) at 60 °C and (b) at 40 °C for 20 hours. (c) and (d) are the ΔA spectra (obtained by subtracting the reference spectrum of the pristine sample from each absorption spectrum).

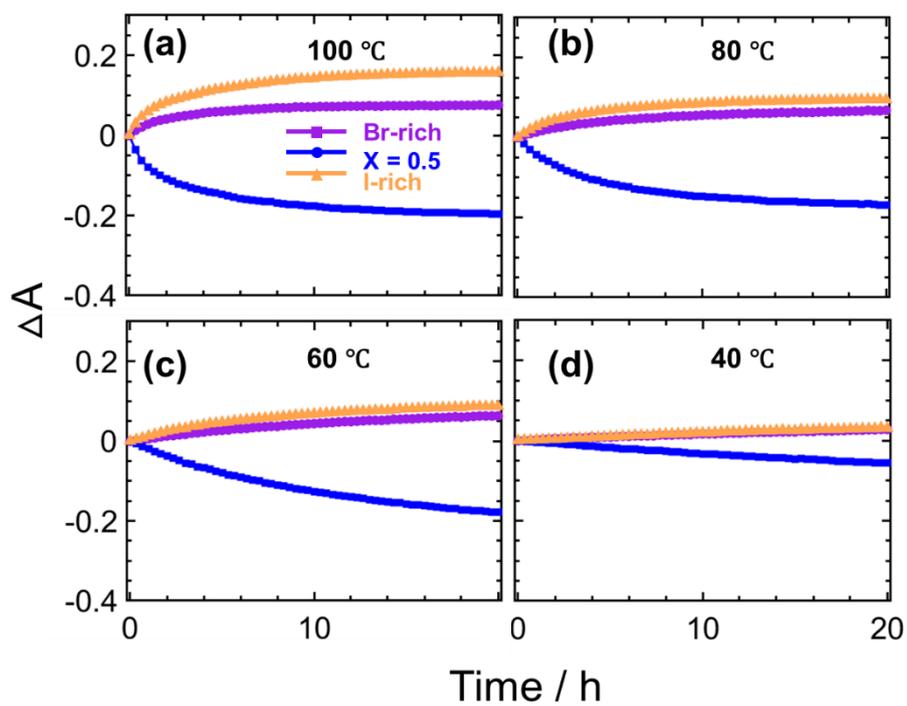

**Figure S15.** Kinetics of photo de-mixing experiments performed at different temperatures. The characteristic wavelengths of the Br-rich, x = 0.5 and I-rich phases are shown. Data extracted from **Figure S13** and **Figure S14**.

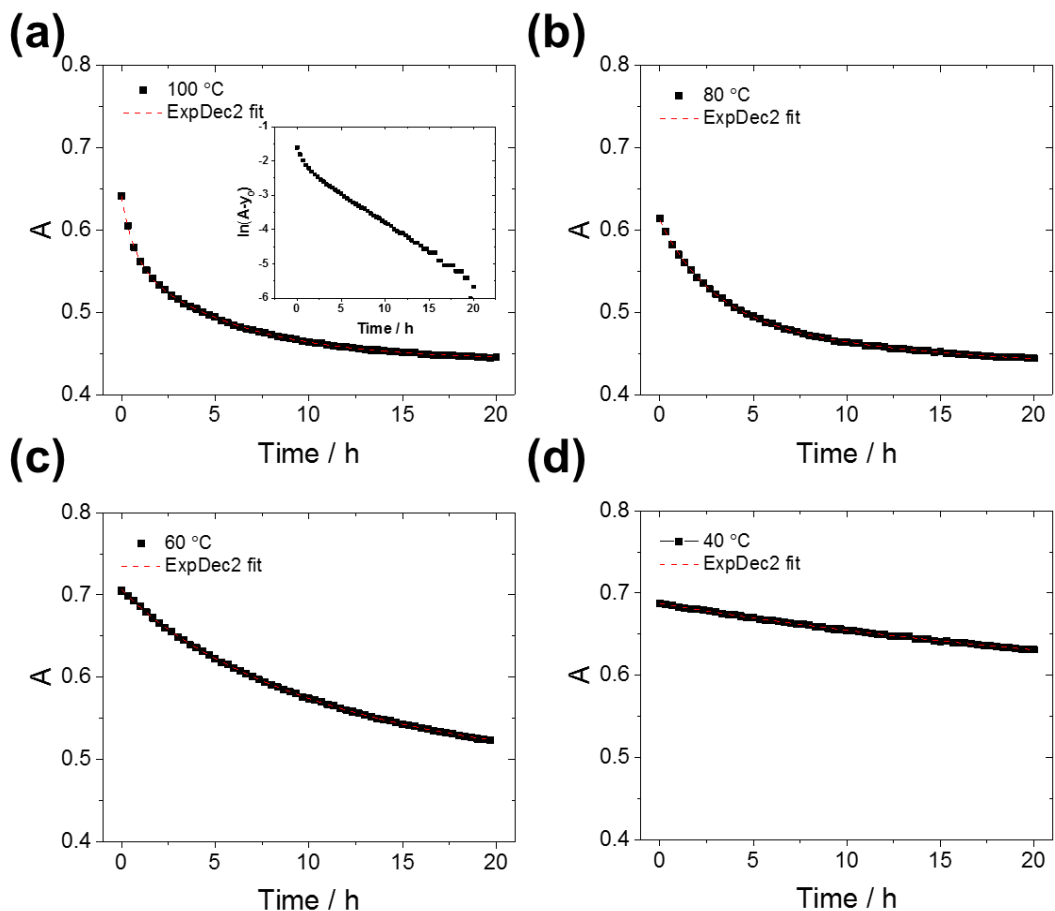

**Figure S16.** (a-c) Kinetics and the fitting for the change in absorbance at the peak absorption wavelength corresponding to the x = 0.5 phase (450 nm) under light at different temperatures. We fitted a bi-exponential to the data (see also inset in (a)). For the fit of the data we apply the constraint that the ratio of the initial absorbance and the absorbance that is reached at steady-state is the same value at different temperatures. Such value is obtained from fitting an unconstrained bi-exponential to the data collected at 100 °C.

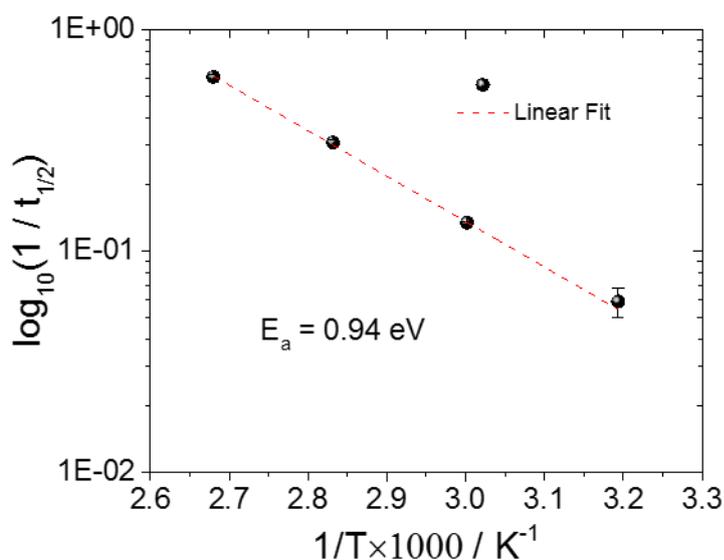

**Figure S17.** Half-time associated to the changes in absorbance shown in **Figure S16** displayed in an Arrhenius plot. The line corresponds to the a linear fit to the data, from which we obtain an activation energy for the photo de-mixing experiment shown in **Figure S12–S16** of $E_{act}$ = 0.94 eV.

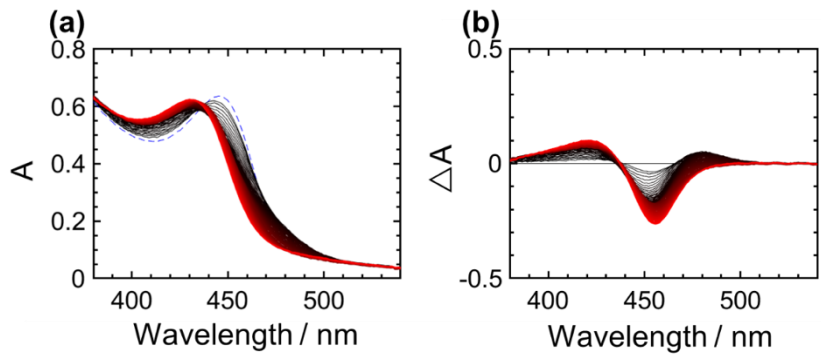

**Figure S18.** (a) UV-Vis absorption spectra evolution of a (PDMA)Pb($I_{0.5}Br_{0.5}$)$_4$ thin film (with PMMA encapsulation) under light (1.5 mW/cm$^2$) at 150 °C (same as the sample annealing temperature for crystallization) for ~40 h. (b) Change in absorbance obtained by subtracting the reference spectrum of the pristine sample from each absorbance spectrum shown in (a). In both graphs, the gradient from black to red denotes the time sequence. In (a) the dashed blue line corresponds to the initial absorbance of the film.

## S7. Photo-induced changes in optical absorption as function of composition

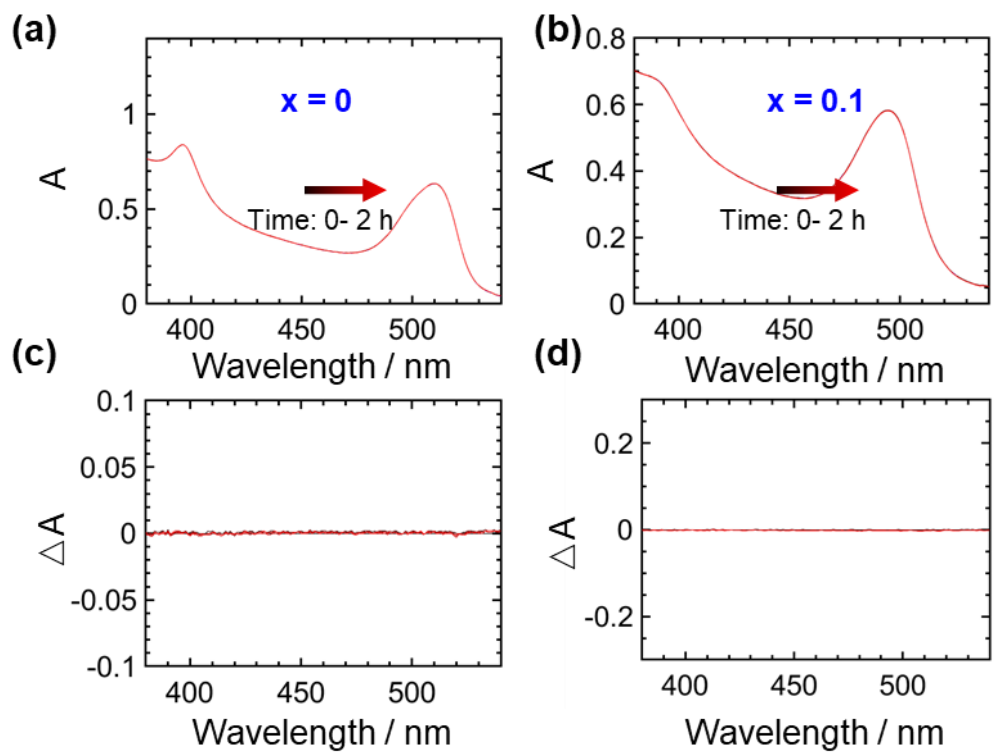

**Figure S19.** UV-vis spectra of (PDMA)PbI$_4$ (a) and (PDMA)Pb(I$_{0.9}$Br$_{0.1}$)$_4$ (b) thin films at 100 °C (with PMMA encapsulation in Ar atmosphere) exposed to 1.5 mW/cm$^2$ illumination for 2 h. No absorption changes detected for both compositions. (c) and (d) are the corresponding ΔA spectra (obtained by subtracting the reference spectrum of the pristine sample from each absorption spectrum).

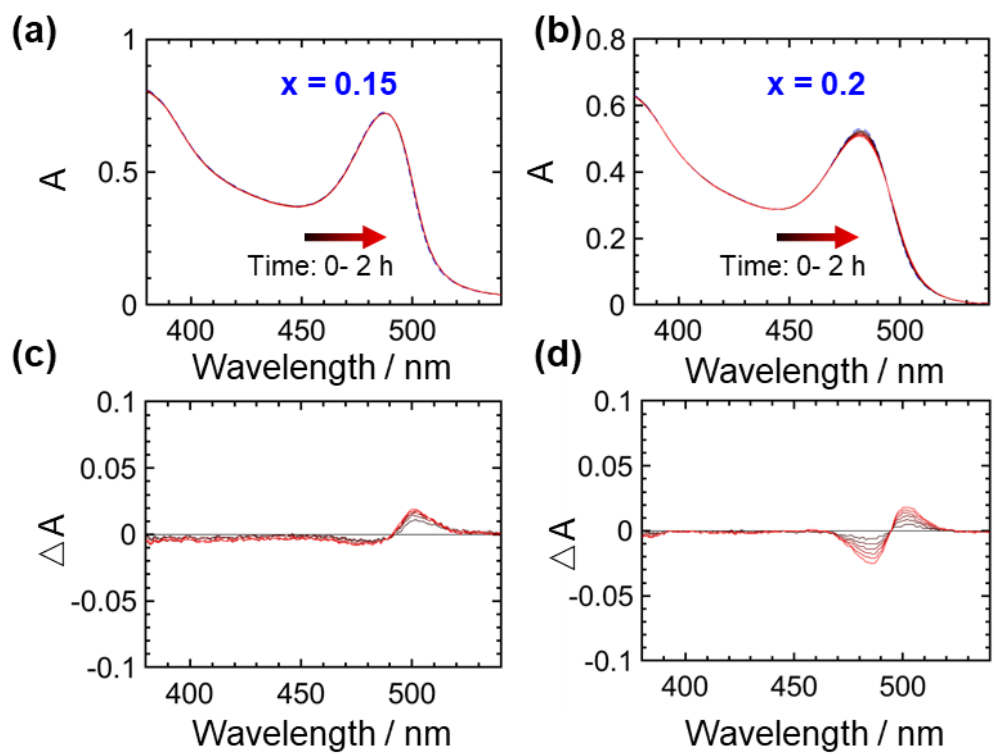

**Figure S20.** UV-vis spectra of (PDMA)Pb(I$_{0.85}$Br$_{0.15}$)$_4$ (a) and (PDMA)Pb(I$_{0.9}$Br$_{0.1}$)$_4$ (b) thin films at 100 °C (with PMMA encapsulation in Ar atmosphere) exposed to 1.5 mW/cm$^2$ illumination for 2 h. (c) and (d) are the corresponding ΔA spectra (obtained by subtracting the reference spectrum of the pristine sample from each absorption spectrum).

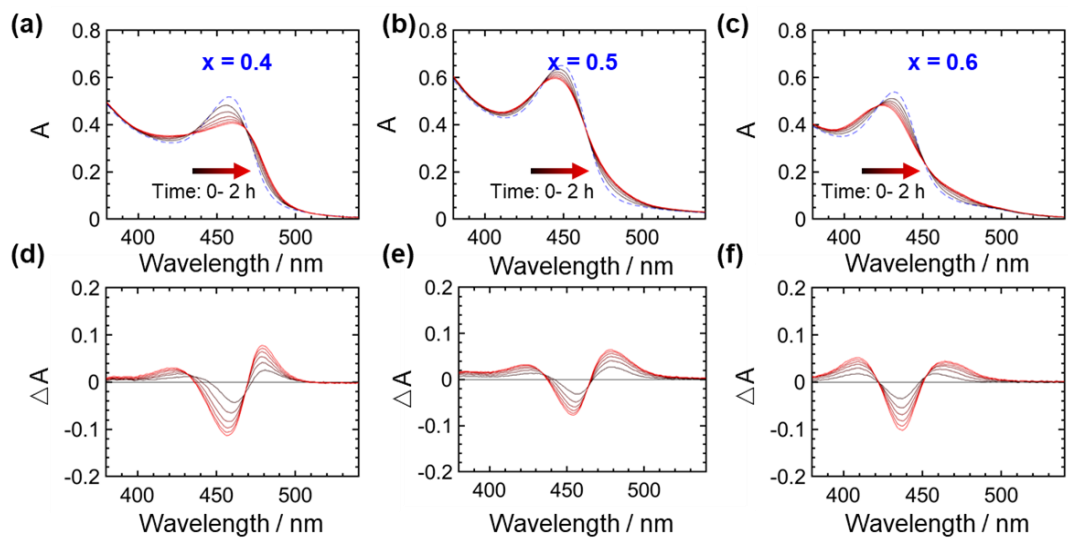

**Figure S21.** UV-vis spectra of (a) (PDMA)Pb(I$_{0.6}$Br$_{0.4}$)$_4$, (b) (PDMA)Pb(I$_{0.5}$Br$_{0.5}$)$_4$ (c) (PDMA)Pb(I$_{0.4}$Br$_{0.6}$)$_4$ thin films at 100 °C (with PMMA encapsulation in Ar atmosphere) exposed to 1.5 mW/cm$^2$ illumination for 2 h. (d-f) are the corresponding ΔA spectra (obtained by subtracting the reference spectrum of the pristine sample from each absorption spectrum).

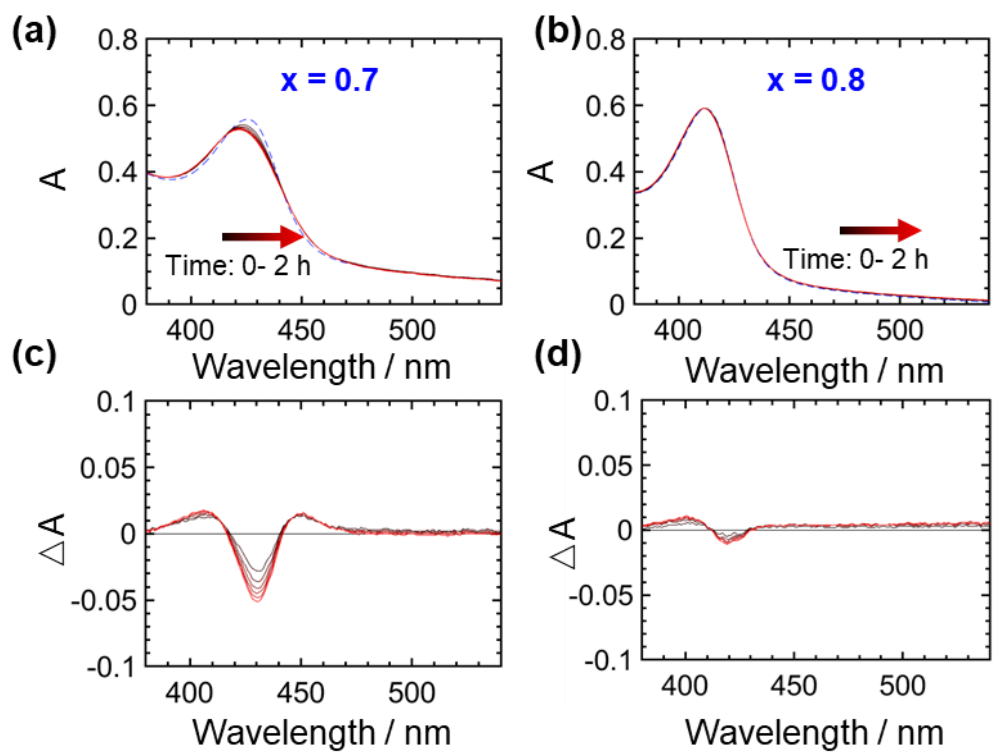

**Figure S22.** UV-vis spectra of (PDMA)Pb(I$_{0.3}$Br$_{0.7}$)$_4$ (a) and (PDMA)Pb(I$_{0.2}$Br$_{0.8}$)$_4$ (b) thin films at 100 °C (with PMMA encapsulation in Ar atmosphere) exposed to 1.5 mW/cm$^2$ illumination for 2 h. (c) and (d) are the corresponding ΔA spectra (obtained by subtracting the reference spectrum of the pristine sample from each absorption spectrum).

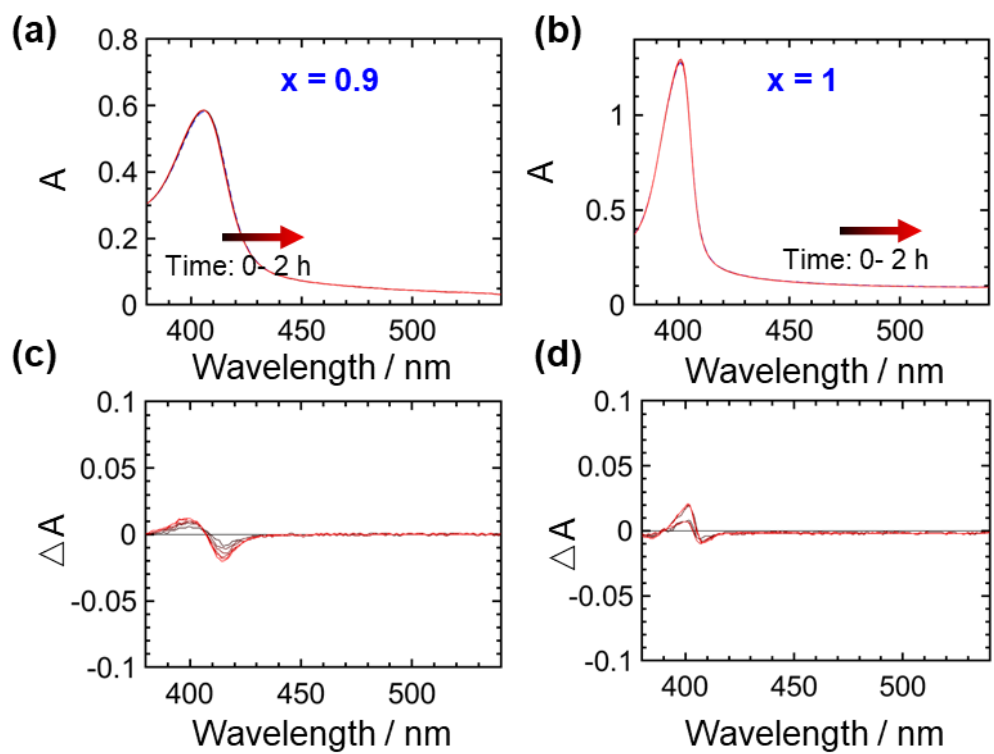

**Figure S23.** UV-vis spectra of (a) (PDMA)Pb($I_{0.1}Br_{0.9}$)$_4$, (b)(PDMA)PbBr$_4$ thin films at 100 °C (with PMMA encapsulation in Ar atmosphere) exposed to 1.5 mW/cm$^2$ illumination for 2 h. (c-d) are the corresponding ΔA spectra (obtained by subtracting the reference spectrum of the pristine sample from each absorption spectrum).

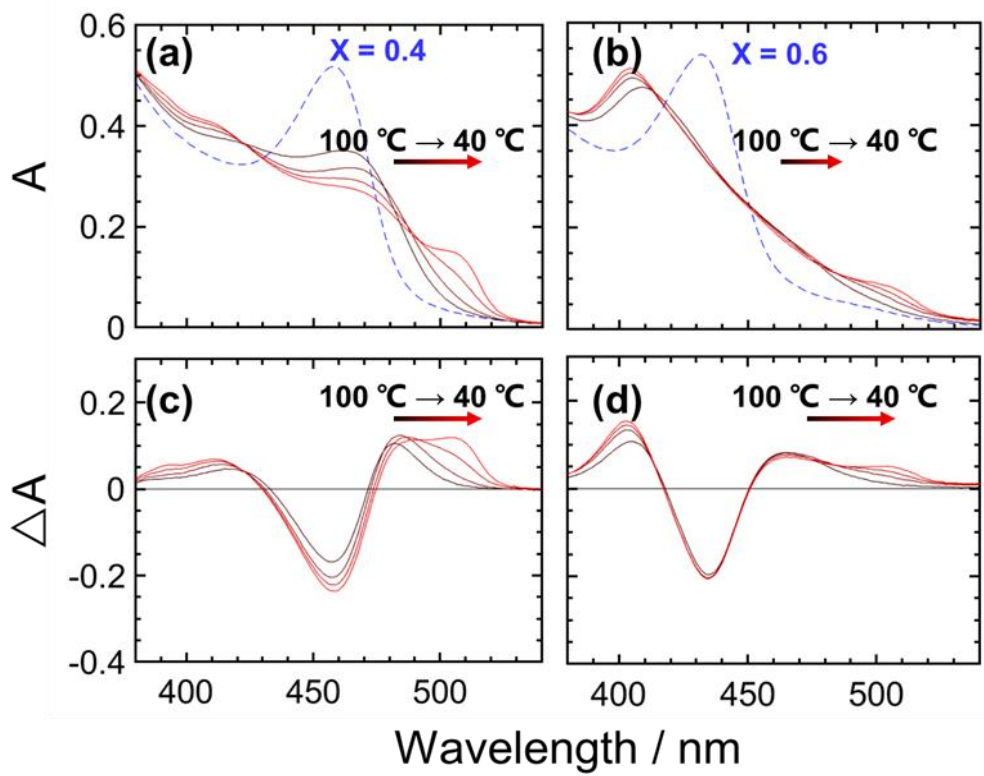

**Figure S24.** Temperature dependence UV-Vis spectra of (a) (PDMA)Pb($I_{0.6}Br_{0.4}$)$_4$ and (b) (PDMA)Pb($I_{0.4}Br_{0.6}$)$_4$ thin films (with PMMA encapsulation in Ar atmosphere) under light (1.5 mW/cm$^2$). The gradient from black to red denotes a decreasing temperature sequence from 100 to 80 to 60 and to 40 °C. Each spectrum corresponds to a measurement performed after the sample was left at the specified temperature for 20 h. The dashed blue line in (a) and (b) denote the absorption spectra of the pristine (PDMA)Pb($I_{0.6}Br_{0.4}$)$_4$ phase ($x_{initial}$ = 0.4) and (PDMA)Pb($I_{0.4}Br_{0.6}$)$_4$ phase ($x_{initial}$ = 0.6), respectively. (c) and (d) are the ΔA spectra (obtained by subtracting the reference spectrum of the pristine sample from each absorption spectrum).

## S8. Light intensity effect on photo-miscibility gap

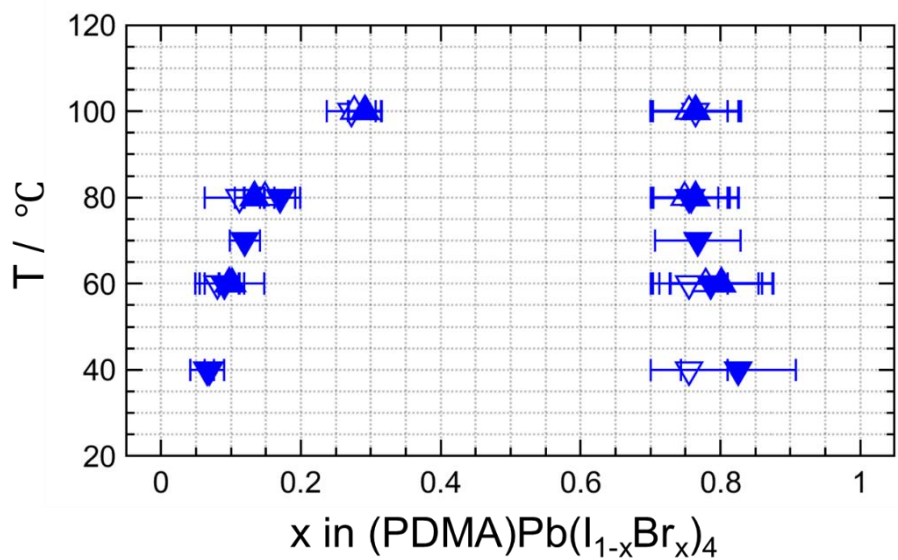

**Figure S25.** Photo-miscibility-gap mapped under ×10 times higher light intensity (15 mW cm$^{-2}$, filled triangles) than the one shown in **Figure 3** (1.5 mW cm$^{-2}$, Empty upwards and downwards pointing triangles). Downward-pointing and upward-pointing triangle symbols denote data collected during a downward and upward temperature scan respectively.

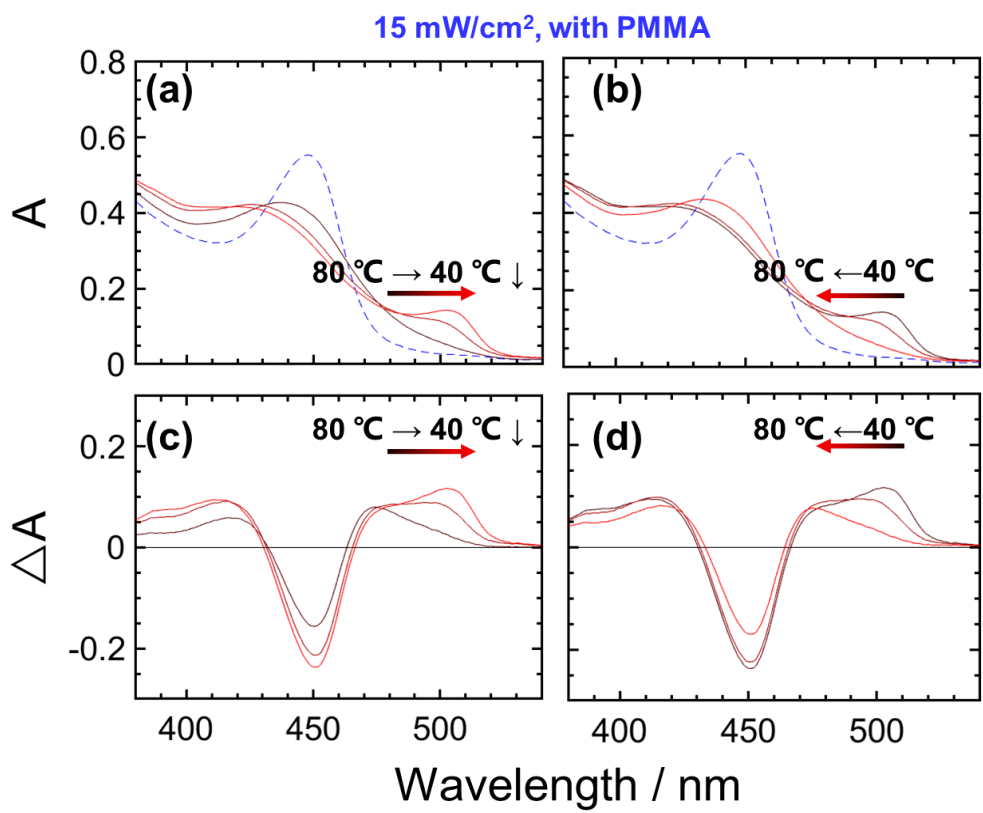

**Figure S26.** Temperature dependence UV-Vis spectra of (PDMA)Pb(I$_{0.5}$Br$_{0.5}$)$_4$ thin film (with PMMA encapsulation) under light (15 mW/cm$^2$) (a) at temperature from 100 °C to 80 to 60 and to 40 °C for 20 h (b) at temperature from 40 °C to 60 to 80 and to 100 °C for 20 h. The gradient from black to red denotes the time sequence of the temperature step. Dashed blue line denotes the absorption spectrum of the pristine (PDMA)Pb(I$_{0.5}$Br$_{0.5}$)$_4$ phase (50:50). (c) and (d) are the ΔA spectra (obtained by subtracting the reference spectrum of the pristine sample from each absorption spectrum).

## S9. Encapsulation effect on photo-miscibility gap

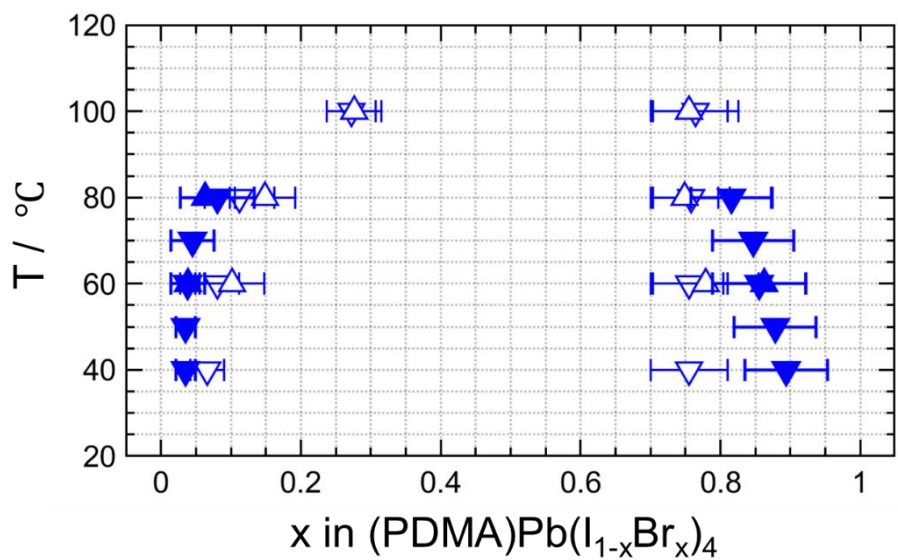

**Figure S27.** Photo-miscibility-gap mapped for a PDMAPb($I_{0.5}Br_{0.5}$)$_4$ thin film without encapsulation (filled triangles). The case of an encapsulated sample as shown in **Figure 3** is also shown for comparison (empty upwards and downwards pointing triangles). Downward-pointing and upward-pointing triangle symbols denote data collected during a downward and upward temperature scan respectively.

## S10. XRD measurements before and after illumination

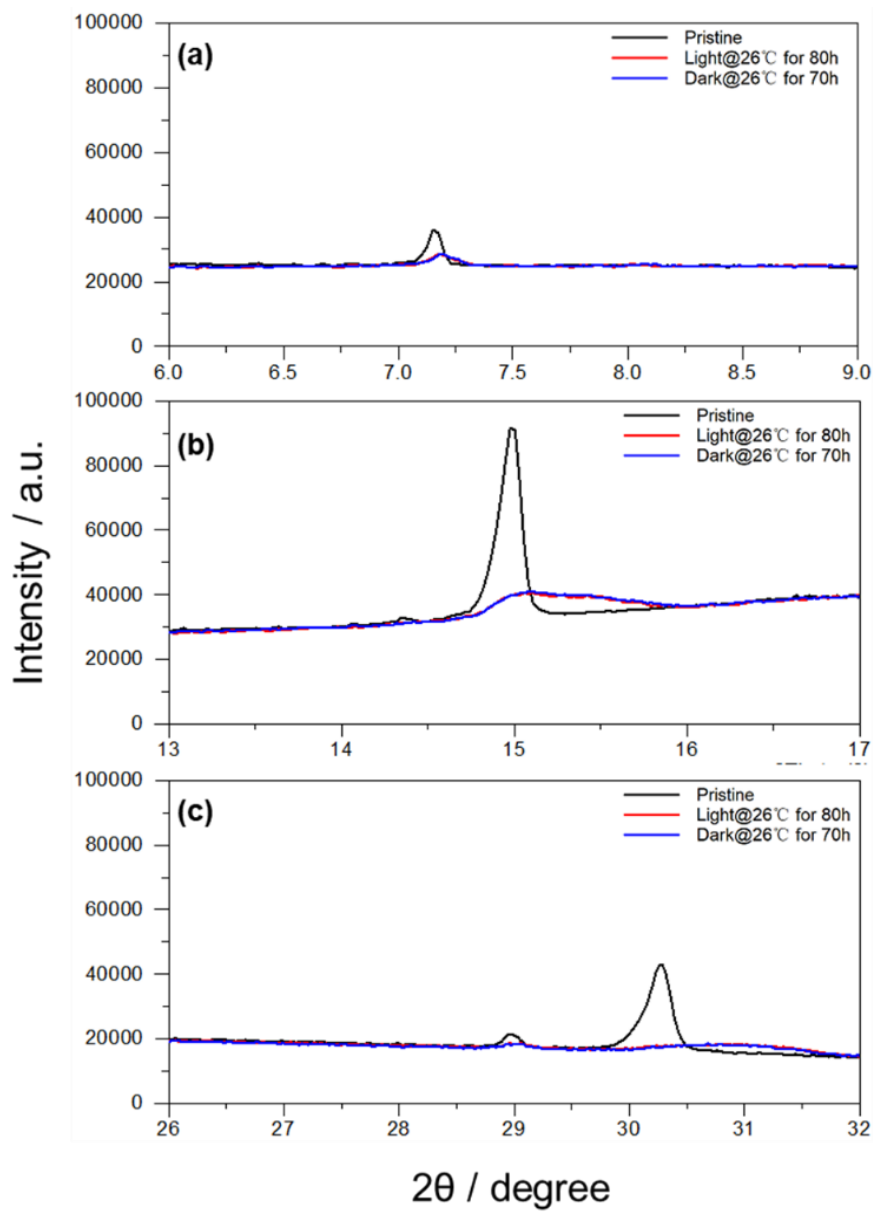

**Figure S28.** XRD measurements performed on a (PDMA)Pb($I_{0.5}Br_{0.5}$)$_4$ thin film (with PMMA encapsulation) highlighting the peaks at (a) $2\theta = 6°- 9°$; (b) $2\theta = 13°- 17°$; (c) $2\theta = 13°- 17°$; before (black) and after illumination at 26 °C for 80 hours (red) and back in the dark at 26 °C for 70 hours (blue).

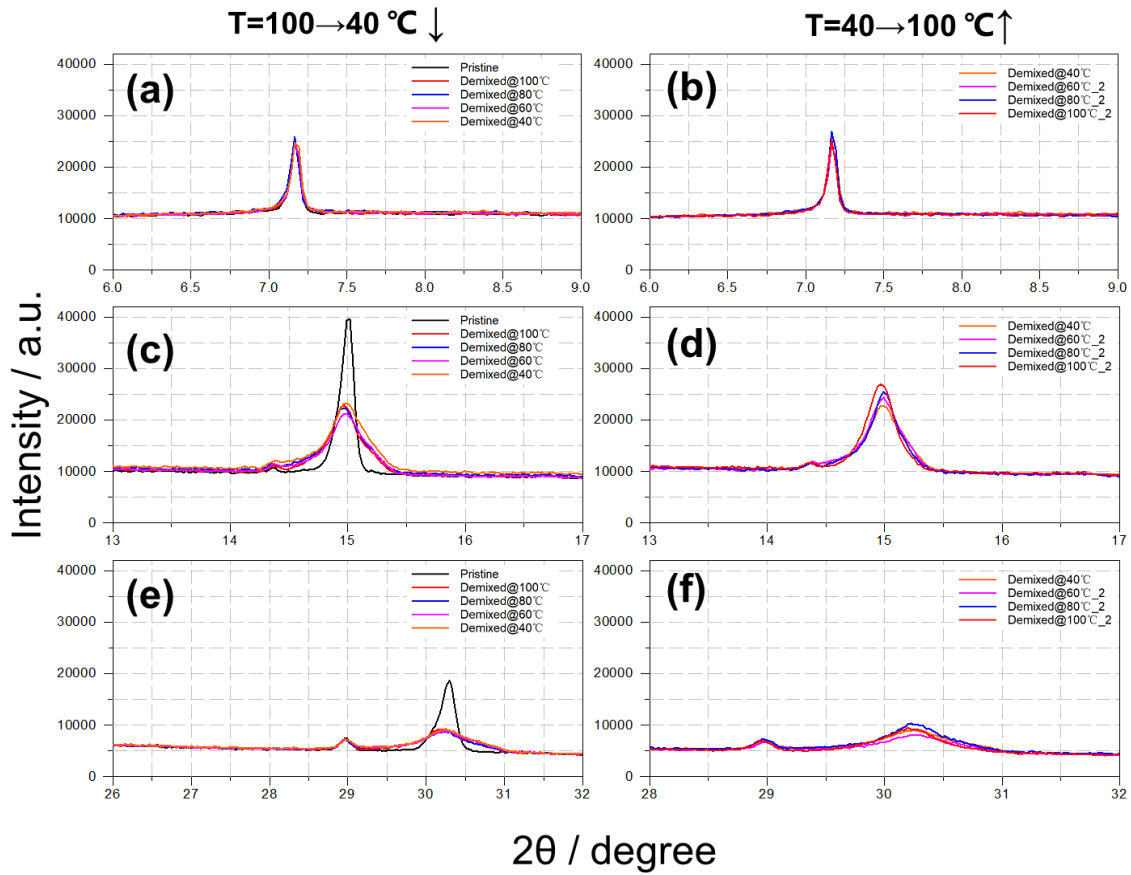

**Figure S29.** XRD measurements performed after each step of the temperature dependent UV-vis study performed on a (PDMA)Pb($I_{0.5}Br_{0.5}$)$_4$ thin film (with PMMA encapsulation) under light (1.5 mW/cm$^2$) (UV-vis data for this sample are shown in **Figure 3** in the main text). The diffraction patterns in different ranges refer to different orientations. (a) 2θ = 6°–9°; (c) 2θ = 13°–17°; (d) 2θ = 13°–17° represent the sample's state at each temperature step after photo de-mixing during the downward temperature scan 100°C → 80°C → 60°C → 40 °C. Diffraction patterns in range (b) 2θ = 6°– 9°; (d) 2θ = 13° – 17°; (f) 2θ = 13° – 17° represent the sample's state at each temperature step after photo de-mixing during the upward temperature scan from 40 °C → 60 °C → 80 °C →100 °C. All XRD measurements were taken at room temperature.

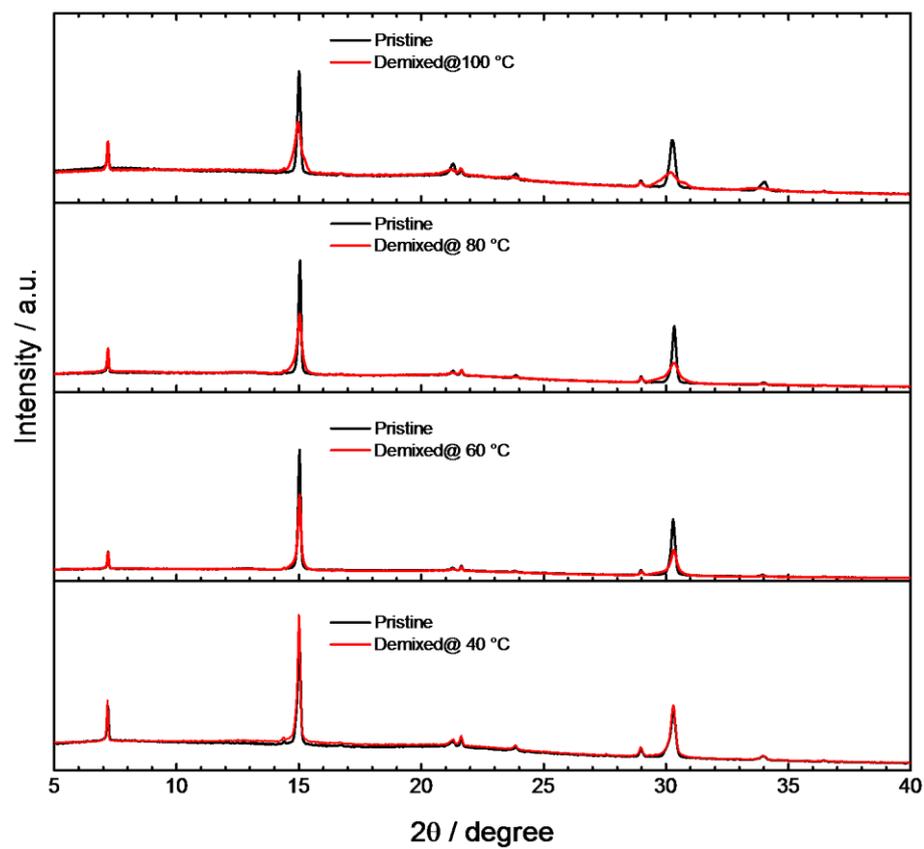

**Figure S30.** XRD measurement for (PDMA)Pb($I_{0.5}Br_{0.5}$)$_4$ thin films (with PMMA encapsulation) after photo de-mixing under light (1.5 mW/cm$^2$) conducted at different temperatures for each sample (UV-Vis study shown in **Figure S12-14**). All XRD measurements were taken at room temperature.

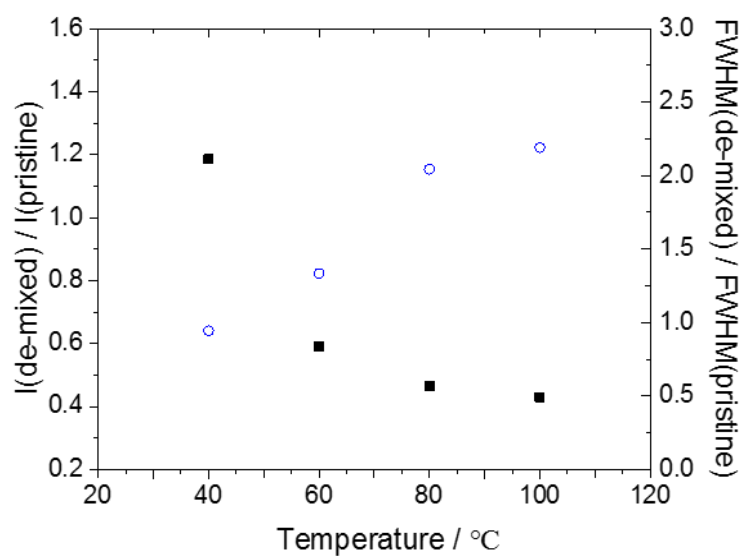

**Figure S31.** Change in relative intensity (black square) and full width half maximum (blue circle) for the XRD measurements shown in **Figure S30** before and after illumination.

## S11 Supporting References